\documentclass[aps,prb,nobibnotes,twocolumn,superscriptaddress,shortbibliography]{revtex4-2}

\usepackage{amsfonts}
\usepackage{mathrsfs}
\usepackage{amsmath}% needed for subequations
\usepackage{color}
\usepackage{graphicx}
\usepackage{bm}% bold maths
\usepackage{amssymb}
\usepackage{xspace}
\usepackage{epstopdf}
\usepackage{dcolumn}% Align table columns on decimal point
\usepackage{longtable}
\usepackage{multirow}
\usepackage{float}
\usepackage{comment}
\usepackage{lineno}
\usepackage[colorlinks=true, letterpaper=true, pdfstartview=FitV,  linkcolor=blue, citecolor=blue, urlcolor=blue]{hyperref}

\makeatother

\begin{document}

\title{Floquet control of topological phases and Hall effects in $Z_2$ nodal line semimetals}

\author{Pu Liu}
\affiliation{School of Microelectronics and Physics, Hunan University of Technology and Business, Changsha 410205, China}
\affiliation{Centre for Quantum Physics, Key Laboratory of Advanced Optoelectronic Quantum Architecture and Measurement (MOE), School of Physics, Beijing Institute of Technology, Beijing, 100081, China}

\author{Chaoxi Cui}
\email{cuichaoxi@bit.edu.cn}
\affiliation{Centre for Quantum Physics, Key Laboratory of Advanced Optoelectronic Quantum Architecture and Measurement (MOE), School of Physics, Beijing Institute of Technology, Beijing, 100081, China}

\author{Lei Li}
\affiliation{Research Center for Quantum Physics and Technologies, Inner Mongolia University, Hohhot 010021, China}
\affiliation{School of Physical Science and Technology, Inner Mongolia University, Hohhot 010021, China}

\author{Runze Li}
\affiliation{Centre for Quantum Physics, Key Laboratory of Advanced Optoelectronic Quantum Architecture and Measurement (MOE), School of Physics, Beijing Institute of Technology, Beijing, 100081, China}

\author{Dong-Hui Xu}
\email{donghuixu@cqu.edu.cn}
\affiliation{Department of Physics, and Chongqing Key Laboratory for Strongly Coupled Physics, Chongqing University, Chongqing 400044, China}
\affiliation{Center of Quantum Materials and Devices, Chongqing University, Chongqing 400044, China}

\author{Zhi-Ming Yu}
\affiliation{Centre for Quantum Physics, Key Laboratory of Advanced Optoelectronic Quantum Architecture and Measurement (MOE), School of Physics, Beijing Institute of Technology, Beijing, 100081, China}
\affiliation{International Center for Quantum Materials, Beijing Institute of Technology, Zhuhai, 519000, China}

\begin{abstract}
Dynamic control of topological properties in materials is central to modern condensed matter physics, and Floquet engineering, utilizing periodic light fields, provides a promising avenue. Here,  we use Floquet theory to theoretically study the topological response of a $Z_2$ nodal line semimetal (NLSM) when driven by circularly polarized light (CPL). We demonstrate that the direction of CPL irradiation critically dictates the resulting topological phase transitions. Specifically, when light is incident perpendicular to the nodal line plane, increasing the light amplitude induces two successive topological phase transitions: first, from the $Z_2$ NLSM to a vortex NLSM, a rare and intriguing topological state; and second, a transition from the vortex NLSM to a semimetal with a pair of Weyl points (WPs). 
In stark contrast, irradiation along other directions directly transforms the $Z_2$ nodal line into a pair of WPs. We further investigate the transport properties of the Floquet $Z_2$ NLSM, focusing on the anomalous and planar Hall effects.
The anomalous Hall effect exhibits a direction-dependent amplitude variation, deviating from conventional two-band NLSM behavior.  Importantly, we reveal a significant and tunable planar Hall effect, a phenomenon largely unexplored in Floquet topological materials, which is highly sensitive to both light amplitude and direction.
Our findings not only present a novel route to realize the vortex NLSM, but also establish an efficient way to control Hall transport phenomena in $Z_2$ NLSMs.
%The planar Hall current (PHC) can be divided into $J^{\mathrm{odd}}$ and $J^{\mathrm{even}}$ components, where $J^{\mathrm{odd}}$ and $J^{\mathrm{even}}$ are the odd and even functions of the relaxation times, respectively.
%When CPL is along the $x$ direction, with magnetic and electric field parallel and perpendicular to the direction of incident light, only $J_x^{\mathrm{even}}$ component exists, and it does not depend on the angle of magnetic field. However, when CPL is along the $z$ direction, both  $J_z^{\mathrm{even}}$ and $J_z^{\mathrm{odd}}$ are nonzero and exhibit a period of $\pi$ by varying the direction of magnetic field. Further, with increasing light intensity, the symmetry of driven system improves, and PHC gradually decreases approaching zero.  
\end{abstract}

\maketitle
\section{Introduction}
Topological semimetals (SMs) represent a novel class of quantum materials characterized by unique band structures where conduction and valence bands intersect in momentum space~\cite{Burkov1,RMPSM,YuSciBull}. This class includes Dirac semimetals (DSMs)~\cite{WangZhijun,WangZhijun2,Madhab,ZKLiu,SMYoung}, Weyl semimetals (WSMs)~\cite{WanXiangang,LvBQ,BQLv2,BurkovAA,YangKaiYu,WengHongming2}, and nodal-line semimetals (NLSMs)~\cite{ChanYH,ChenFang,YuanpingChen,Soluyanov,Burkov}, all of which have attracted significant attention in recent years due to their unique band structures and exotic quasiparticles. DSMs exist in systems preserving both time-reversal $\mathcal{T}$ and spatial inversion $\mathcal{P}$ symmetries, resulting in four-fold degenerate band crossings at discrete points. In contrast, WSMs require the breaking of either $\mathcal{T}$ or $\mathcal{P}$ symmetry, leading to pairs of Weyl points (WPs) with opposite chirality~\cite{WanXiangang,LvBQ,BQLv2,BurkovAA,YangKaiYu}. In contrast to DSMs or WSMs, where the conduction and valence bands intersect at discrete points in momentum space~\cite{RMPSM,SMYoung,WanXiangang}, NLSMs feature continuous lines or loops of band crossing, giving rise to distinct topological and transport phenomena~\cite{WengHongming,Burkov,XMZhang,FangChen2,Chiu,Youngkuk}. 

Among NLSMs, conventional NLSMs are topologically protected by a $\pi$ Berry phase and stabilized by symmetries such as $\mathcal{PT}$ or mirror symmetries~\cite{ChanYH,FangChen2,Chiu,ZhaoYX}. However, theoretical advances have identified a special variant: the $Z_2$ NLSM~\cite{SongZhida,FangChen2,Kawakami,Ahn}, distinguished by a nontrivial $Z_2$ monopole charge originating from the topology of real electronic states~\cite{Morimoto,ZhaoYX2,FangChen2,Ahn,ChenCong}. Nodal lines~(NLs) with $Z_2$ monopole charges can only be created and annihilated in pairs. $Z_2$ NLs can appear as a consequence of double band inversion~\cite{Ahn,FangChen2,Kawakami}. 
This distinctive band topology has been predicted in electronic band structures~\cite{Ahn,Kawakami,ChenCong,XGAO,Nomura}, phonon spectra~\cite{Yilin,XTwang}, artificial systems~\cite{YongpanLi,Haoran,PWu}, and even as a photoinduced phase in $Z_2$ DSMs under periodic driving~\cite{Salerno}.
%ABC-stacked graphdiyne~\cite{Ahn,Kawakami,ChenCong,XGAO,Nomura} and 1T'-MoTe$_2$~\cite{WangZhijun3}, and 
Additionally, another unconventional type, the vortex NLSM, has been proposed in three-dimensional systems with mirror symmetry~\cite{Tomas,Lenggenhager}. Protected by both a $\pi$ Berry phase and a $Z$ charge defined on a hemisphere enclosing the NL, vortex NLSMs are rare, with experimental realization only recently achieved in circuit metamaterials~\cite{WuMaopeng} and  proposals in strained ZrTe and Li$_2$NaN~\cite{Tomas2,Lenggenhager}.

Floquet engineering, the use of periodic external fields such as light to dynamically manipulate material properties, has emerged as a powerful tool for tailing topological phases of matter~\cite{MarkS,Takashi2,ChanghuaBao,YHWANG,Fahad,Edbert,YHWANG2}. A typical example of this is the realization of the Floquet topological insulator~\cite{Netanel,Titum,Titum2}, demonstrating the ability of periodic light fields to create nontrivial band topology. Specifically, circularly polarized light (CPL) has been predicted to induce a gap at the degenerate Dirac points of graphene, leading to a quantum Hall effect even without a magnetic field~\cite{Oka,FuLiang,Usaj}. Periodic optical fields have also been used to generate WSMs from various parent states, including topological insulators, DSMs, and NLSMs \cite{Chan,ChenRui,Yan2,DengTingwei,Hannes,LiuHang,YanZhongbo2,Awa}. Recent proposals even suggest the possibility of photoinduced higher-order topological insulators~\cite{Ghosh,Rodriguez,Huang,ZhuWeiwei} and higher-order topological semimetals~\cite{DuXiuLi,ZMWang,BQWang,GhoshWSM}. Beyond these topological phases, periodically driven systems can exhibit various novel transport phenomena, such as a photoinduced tunable anomalous Hall effect~\cite{DHXu,Chan2,Amit,Haowei,DHXu4,YanZhongbo}, as well as thermal Hall and Nernst effect~\cite{Menon,Nag}. Notably, Floquet engineering opens avenues to generate topological states and physical properties that are absent in equilibrium conditions~\cite{Castro,ZhuWeiwei,Higashikawa,LiLinhu}. In the context of NLSMs, this technique is particularly promising for tailoring their anisotropic band structures and rich topological responses. %Therefore, the periodically driven external field provides an efficient and powerful avenue for exploring and manipulating topological states and properties.

%Recently, Du et. al.~\cite{DuXiuLi} investigated the Floquet states of the $Z_2$ NL via lattice models, focusing on the higher-order topology and topological boundary states. 
In this work, we employ a $k\cdot p$ model combined with Floquet theory to theoretically investigate the topological transitions and transport properties of a $Z_2$ NLSM driven by CPL.
Unlike conventional NLSMs, the $Z_2$ NLSM originates from a double band inversion and exhibits a unique linking structure, requiring a four-band model for its description. 
Consequently, the behavior of the $Z_2$ NL is fundamentally different from that of a normal NL.
Due to the anisotropic band structure of the $Z_2$ NL, its response to CPL depends strongly on the light’s propagation direction. Assuming the $Z_2$ NL resides at the $k_x$=0 plane, 
irradiation with the CPL along $x$ direction induces two successive topological transitions as the light amplitude increases: first from the $Z_2$ NL to a vortex NL, then to a pair of WPs located on the $k_x$ axis. The separation between these two WPs increases with increasing amplitude of light. In contrast, when the CPL is incident along the $y$ or $z$ direction, the $Z_2$ NL undergoes a direct transformation into a pair of WPs located on the $k_y$ or $k_z$ axis, respectively. 

The presence of CPL induces a finite photo-induced anomalous Hall conductivity~(AHC) in the $Z_2$ NLSM.
At low amplitudes of light, the AHC exhibits a nonmonotonic dependence on the amplitude. 
However, when the amplitude of light is sufficiently strong, or the Fermi energy of the system is located near the WPs, the AHC is directly proportion to the square of the amplitude of light.
We further study the  planar Hall conductivity~(PHC) in the driven systems, and found that the PHC can also be tuned by adjusting the direction and amplitude of the incident light.
The PHC can be divided into odd $J^{\mathrm{odd}}$ and even components $J^{\mathrm{even}}$ with respect to the relaxation time. 
For simplicity, we focus on the scenario where external magnetic and electric fields are aligned. When the CPL is along $x$ direction, the PHC consists only of $J_x^{\mathrm{odd}}$ component, remains independent of the magnetic field angle. 
In sharp contrast, for the system driven by CPL along the $z$ direction, both $J_z^{\mathrm{odd}}$ and $J_z^{\mathrm{even}}$ components exhibit a periodicity of $\pi$ as the direction of fields is varied. 
%Further, regardless of the direction of incident light, the symmetry of driven system improves as light intensity increases, leading to a gradual decrease in PHC towards zero. 

This paper is organized as follows. In Sec. \ref{sec II}, we derive the effective Floquet Hamiltonian of the Z$_2$ NLSM under off-resonant CPL using a four-band $k\cdot p$ model and Floquet theory, supplemented by two-band models near the $\Gamma$ point and the photo-induced WPs. In Secs. \ref{sec III} and \ref{sec IV}, we respectively study the AHC and PHC under the irradiation of CPL in different directions. Conclusions are given in Sec. \ref{sec:V}.

\section{Floquet Hamiltonian of the periodically driven $Z_2$ NLSM}\label{sec II}
\subsection{General description}
\begin{figure}
\includegraphics[width=1\columnwidth]{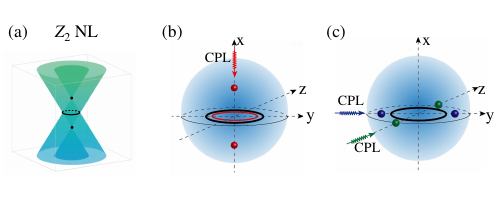}
\caption
{(a) A $Z_2$ NL has band-touching rings on the $k_x$=0 plane. (b) Schematic diagram of topological phase transitions for $Z_2$ NL driven by CPL in $x$ direction. The red ring (denoting vortex NL) and dots (denoting WPs) represent the case for the amplitude of light $A<A_c$ and $A>A_c$, respectively. (c) Schematic diagram of topological phase transitions of $Z_2$ NL driven by CPL in $y$ and $z$ direction, which represent by the blue and green dots (denoting WPs), respectively. 
The black rings in (b-c) denote the original $Z_2$  NL without CPL.}
\label{Fig.1}
\end{figure}

$Z_2$ NLSM that arises from a double band inversion is generally described by a four-band model~\cite{FangChen2,Ahn},  
\begin{eqnarray}
H_0(\bm{k})=k_x \sigma_x +k_y \tau_y \sigma_y+k_z\sigma_z+m \tau_z \sigma_z,\label{Eq.1}
\end{eqnarray}
where $\tau_{x,y,z}$ and $\sigma_{x,y,z}$ are Pauli matrices acting on two isospin degrees. The energy dispersion of the system is
\begin{eqnarray}
\varepsilon(\bm{k})=\pm\sqrt{k_x^2+(k_{||}\pm m)^2}, 
\end{eqnarray}
where $k_{||}=\sqrt{k_y^2+k_z^2}$. 
For half filling, the conduction and valence bands cross and form a  closed loop, which is located on the $k_x$=0 plane with a radius of $|m|$. Besides, the two occupied bands cross along another line  $k_{||}$=0, which is linked with the $Z_2$ NL. The schematic band structure of this model is shown in Fig.~\ref{Fig.1}(a), with $Z_2$ NL plotted by the black ring and the other band degeneracy points represented by black dots.

To explore the impact of periodic driving on the band structure and topological response of the $Z_2$ NLSM, we subject the system to a periodic CPL applied in three different directions. The period of CPL is $T=\frac{2\pi}{\omega}$ with $\omega$ being the frequency of light. The intensity of CPL can be expressed as  
$\bm{{\cal{E}}}(\tau)=-\partial_{\tau} \bm{A}(\tau)$ with a time periodic vector potential $\bm{A}(\tau)=\bm{A}(\tau+T)$.
The Hamiltonian and eigenstates of the driven system become periodic and fulfils $H(\tau)=H(\tau+T)$, and $|\Phi(\tau)\rangle=|\Phi\rangle(\tau+T)$.
Floquet theory simplifies the solution of the Schrödinger equation for a time periodic Hamiltonian to an eigenvalue problem for a time-independent~\cite{MarkS,Takashi2}.
Using the Fourier transformation $H(\tau)=\sum_n e^{-in\omega\tau}H_n$ and 
$|\Phi(\tau)\rangle=\sum_n e^{-in\omega\tau}|\Phi^n\rangle$, the time-dependent Schrödinger equation is mapped to a static eigenvalues problem~\cite{JonHS,HideoS},
\begin{eqnarray}
\sum_n (H_{m-n}-n\omega\delta_{mn})|\Phi_{\alpha}^n\rangle=\epsilon_{\alpha}|\Phi_{\alpha}^n\rangle,
\end{eqnarray}
in the extended Floquet space. 
%We are concerned with the transitions of electrons through the absorption or emission of photons, which always occurs in the case of high-frequency light driving.
Here, we are concerned with the high-frequency light, whose frequency is much larger than the energy scales of the system.
According to the Magnus expansion, the effective Floquet Hamiltonian up to $1/\omega$ can be expanded as~\cite{Oka,FuLiang,Usaj}
\begin{eqnarray}
H_{\mathrm{eff}}=H_0+\frac{[H_{-1},H_1]}{\omega},
\end{eqnarray}
where the higher-order terms of $1/\omega$ is dropped.
Due to the anisotropic band structure of the $Z_2$ NL, we investigate its topological phase transitions under driving by the CPL along different directions. 

\subsection{CPL along $x$ direction}
We first study the system driven by CPL along $x$ direction, with vector potential $\bm{A}=A(0,\eta \sin\omega\tau,\cos\omega\tau)$. Since amplitude of light $A=e{\cal{E}}\omega^{-1}$ with ${\cal{E}}$ denoting field strength, it is adjustable in the experiment by simply tuning ${\cal{E}}$. $\eta=1$ and $\eta=-1$ represent the right- and left-handed CPL, respectively.
Following the standard approach, the electromagnetic coupling is given by $H(\bm{k})\rightarrow H[\bm{k}-e\bm{A}(\tau)]$. In this case, the effective Floquet $k\cdot p$ Hamiltonian in the high frequency limit becomes
\begin{eqnarray}
H_{\mathrm{eff}}^{x}=H_0-\frac{1}{2\omega}i\eta A^2[\sigma_z,\tau_y\sigma_y].
\label{Eq.5}
\end{eqnarray}
Compared to undriven system Eq.~(\ref{Eq.1}), the Hamiltonian gains an additional correction term. Although the $\mathcal{T}$ symmetry of the driven system is broken by CPL, it still has 
$\mathcal{C}_{2y(2z)}\mathcal{T}$ symmetry. 
The eigenvalues for the driven system is obtained as
\begin{eqnarray}\label{Eq.6}
\varepsilon&=&\pm\sqrt{k_{||}^2+k_x^2+m^2+\frac{A^4}{\omega^2}\pm 2\sqrt{m^2k_{||}^2+\frac{A^4}{w^2}(k_x^2+m^2)}}.\notag \\
\end{eqnarray}

\begin{figure}
\includegraphics[width=1\columnwidth]{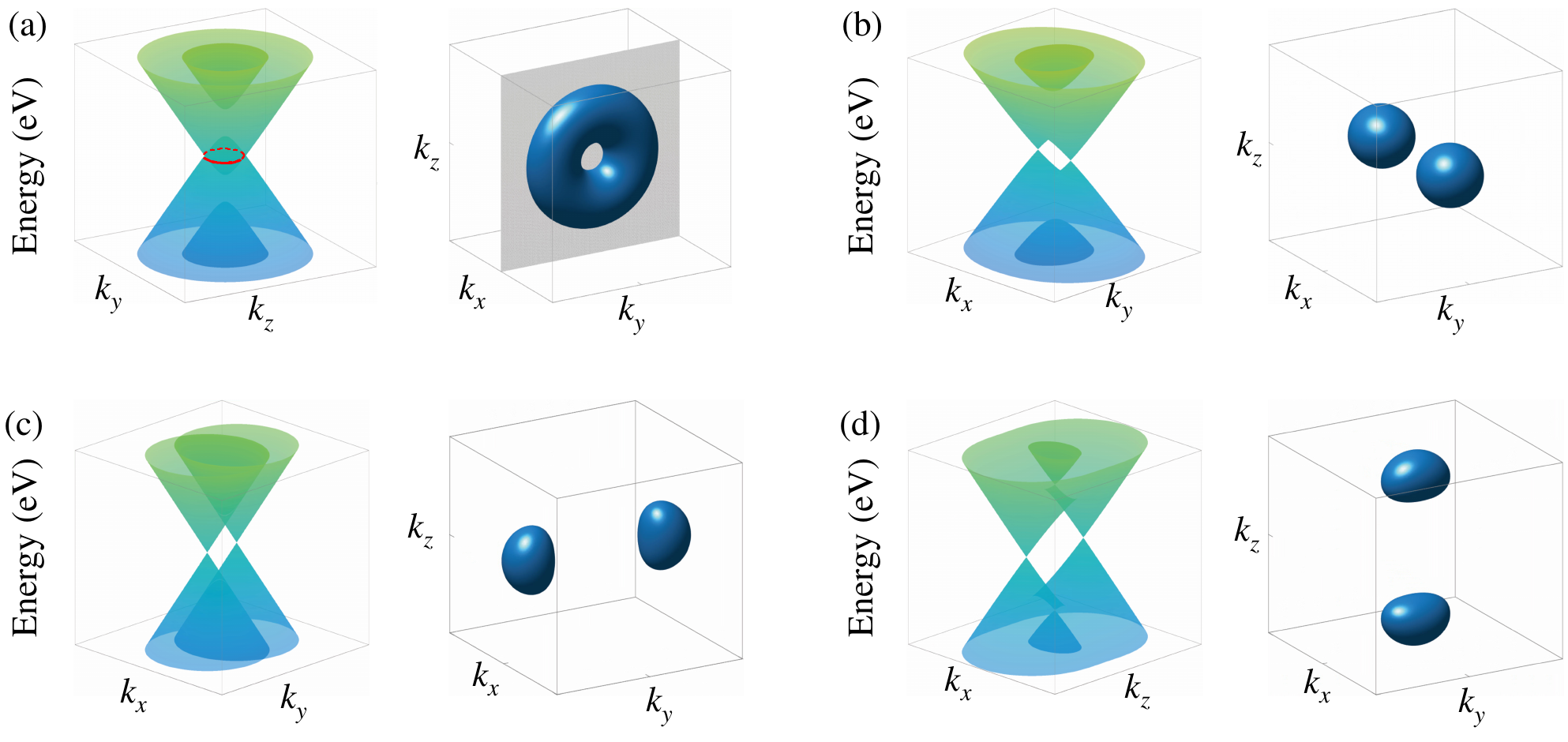}
\caption
{The band structures and Fermi surfaces for $E_F=0.05$ eV for $Z_2$ NL driven by CPL. The CPL is along $x$ direction with amplitudes of light (a) $A$=0.5 $\mathrm{\AA}^{-1}<A_c$, and (b) $A$=1 $\mathrm{\AA}^{-1}>A_c$. The  CPL is along (c) $y$ direction and (d) $z$ direction with $A$= 1 $\mathrm{\AA}^{-1}$. We take the other model parameters as $m$=0.1 eV and $\omega$=6 eV.}
\label{Fig.2}
\end{figure}

\begin{figure}
\includegraphics[width=1\columnwidth]{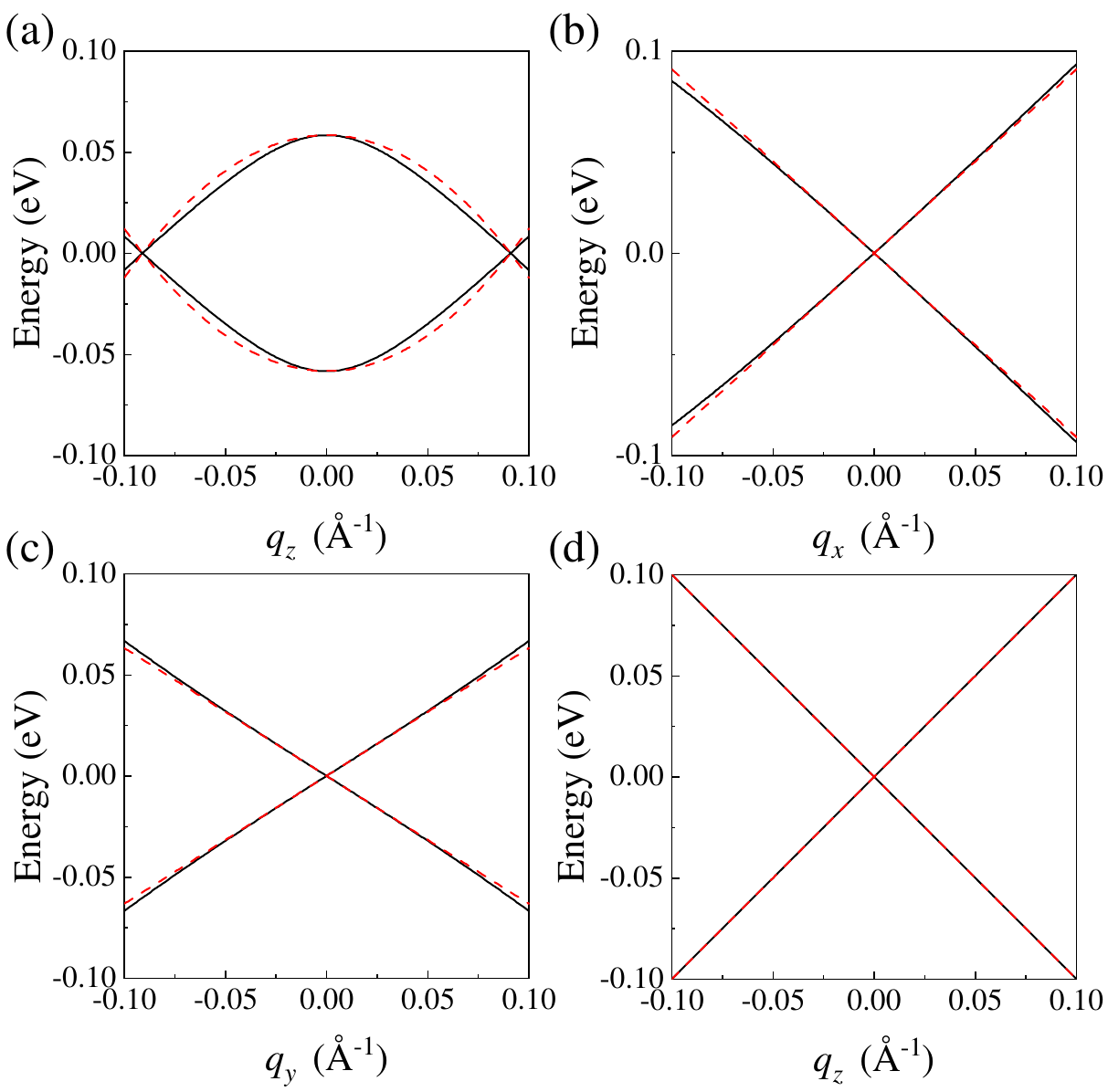}
\caption
{The band structure for driven systems around $\Gamma$ and WP. (a-b) The CPL is along $x$ direction with amplitudes of light $A$=0.5 \AA$^{-1}$ and $A$=1.2 \AA$^{-1}$, respectively. (c-d) The CPL is along $z$ direction with amplitude of light $A$=0.7 \AA$^{-1}$. The band plotted by the black line are calculated by the Eq.~\ref{Eq.5} and Eq.~\ref{Eq.9}, which plotted by the red dashed line are calculated by the two band model Eq.~\ref{Eq.7}, Eq.~\ref{Eq.8} and Eq.~\ref{Eq.10}. We take the model parameters as $m$=0.1 eV and $\omega$=6 eV.}
\label{Fig.5}
\end{figure}

The evolution of band structures with increasing $A$ is schematically depicted in  Fig.~\ref{Fig.1}(b). The black ring on the $k_x$=0 plane represents the $Z_2$ NL for the original system. As $A$ increases, the driven system undergoes two topological phase transitions. In the regime of low light amplitude, specifically when $A<A_c$ ($A_c=\sqrt{m\omega}$), the radius of the NL located on the $k_x$=0 plane continuously decreases as  $A$ increases, which is plotted as the red ring in Fig.~\ref{Fig.1}(b). Simultaneously, the degeneracy between the two  occupied  states is lifted, breaking the  linking structure of the $Z_2$ NL [see Fig.~\ref{Fig.2}(a)].
Intriguingly, the resulting NL is not a conventional NL but a vortex NL.
To directly demonstrate it, we derive a two-band effective  model  from Eq.~(\ref{Eq.5}) to  describe the low-energy bands, expressed as
\begin{eqnarray}
\begin{aligned}
H_{vNL}=&\frac{1}{m+\frac{A^2}{\omega}}[(k_x^2-k_y^2-k_z^2+m^2-4\frac{A^4}{\omega^2})\sigma_z\\&+2k_xk_y\sigma_x+2ik_xk_z\sigma_y].
\label{Eq.7}
\end{aligned}
\end{eqnarray}
Using the two-band model, we plotted the band structure (red dashed line) near the $\Gamma$
point in Fig.~\ref{Fig.5}(a) and compared it with the result calculated from Eq.~\ref{Eq.5} (black solid line). This comparison validates the accuracy of our effective two-band description near the zone center. To further characterize the topological nature of the resulting NL, we calculated the winding number of both upper~($k_x>0$ region) and lower~($k_x<0$ region) hemispheres on each side of the NL~\cite{Tomas}. These calculations yielded winding numbers of $w=+1$ and $w=-1$, respectively. This is distinct from the normal NL and the $Z_2$ NL, both of which exhibit a winding number of $w=0$. The presence of non-zero and opposite winding numbers classifies this specific type of NL as a vortex~\cite{Tomas,Lenggenhager}. Therefore, our findings indicate that  under driving of CPL in $x$ direction with an amplitude below a critical value $A<A_c$, the $Z_2$ NL undergoes a transition to a vortex NL, thus presenting a potentially straightforward method for the experimental realization of this intriguing topological state.

At the critical point $A=A_c$, the vortex NL on the $k_x=0$ plane shrinks into a point. Further increasing $A$, a pair of WPs appear, residing on the $k_x$ axis, as shown in the Fig.~\ref{Fig.1}(b) and Fig.~\ref{Fig.2}(b). The locations of the WPs are tunable by changing amplitude and frequency of CPL , which is ${\bm k}_{w,x}^{\pm}=\pm(\sqrt{\frac{A^4}{\omega^2}-m^2},0,0)$. 
We also establish an effective Hamiltonian to investigate the topological properties  of these WPs. 
Up to the linear term,  the effective two-band  Hamiltonian is obtained as 
\begin{eqnarray}
\begin{aligned}
H^{x}_{\pm}=&-\sqrt{\frac{ A^4-\omega^2 m^2}{A^4}}(\mp q_x\sigma_z+q_z\sigma_x+q_y\sigma_y),
\end{aligned}
\label{Eq.8}
\end{eqnarray}
where the momentum ${\bm q}$ is measured from $\pm(\sqrt{\frac{A^4}{\omega^2}-m^2},0,0)$. 
The band structure near a WP at ${\bm k}_{w,x}^{+}$  with $A=1.2$ \AA$^{-1}$ is shown in Fig.~\ref{Fig.5}(b), where the red dashed lines are calculated by the Eq.~(\ref{Eq.8}), and the black solid lines are calculated by Eq.~(\ref{Eq.5}).
One observes that the band structures are almost the same around the WP.
Furthermore, we calculate the winding number (Chern number) of the WP via the effective Hamiltonian (\ref{Eq.8}), and find that the two WPs at ${\bm k}_{w,x}^{\pm}$ have opposite winding number of $w=\pm 1$.
%By increasing $A$, the effect of $h^x_{\mathrm{other}}$ diminishes. When light intensity $A$ is very large, the linear terms become the leading order in Eq.~\ref{Eq.8}, which makes the WP almost an ideal WP, and the driven system also has high symmetry. 

\subsection{CPL along $z$  direction}
When the propagation direction of CPL is along the $z$ direction $\bm{A}=A(\eta \sin\omega\tau,\cos\omega\tau, 0)$, the effective Floquet Hamiltonian is established as 
\begin{eqnarray}
H_{\mathrm{eff}}^{z}=H_0-\frac{1}{2\omega}i\eta A^2[\sigma_x,\tau_y\sigma_y],
\label{Eq.9}
\end{eqnarray}
for which the  eigenvalue is 
\begin{eqnarray}
\varepsilon=\pm\sqrt{k_{||}^2+k_x^2+m^2+\frac{A^4}{\omega^2}\pm 2\sqrt{m^2k_{||}^2+\frac{A^4}{w^2}k_z^2}}. \notag \\
\end{eqnarray}
When CPL is incident along the $z$ direction, the $Z_2$ NL will directly transform into a pair of WPs along the $z$ axis located at ${\bm k}_{w,z}^{\pm}=(0, 0, \pm\sqrt{\frac{A^4}{\omega^2}+m^2})$, as illustrated in  Fig.~\ref{Fig.1}(c). 

Similarly, if CPL is along the $y$ direction, the Floquet Hamiltonian becomes 
\begin{eqnarray}
H_{\mathrm{eff}}^{y}=H_0-\frac{1}{2\omega}i\eta A^2[\sigma_x,\sigma_z],
\end{eqnarray}
where two  WPs located at ($0, \pm\sqrt{\frac{A^4}{\omega^2}+m^2}, 0$) appear [see the blue dots in Fig.~\ref{Fig.1}(c)].  The locations for the WPs can also be adjusted by amplitude  and frequency of CPL. Moreover, as shown in Fig.~\ref{Fig.2}(c-d), the band structures of the system driven by CPL in $y$ or $z$ direction are similar. Therefore, in subsequent calculations, we only show the result for the system driven by CPL in $z$ direction.

We also obtained a two-band model to describe the WPs located on $k_z$ axis, expressed as  
\begin{eqnarray}
\begin{aligned}
H^{z}_{\pm}=q_z\sigma_z+q_x\sigma_x\pm\sqrt{\frac{A^4}{m^2w^2+A^4}}q_y\sigma_y,
\label{Eq.10}
\end{aligned}
\end{eqnarray}
where the momentum ${\bm q}$ is measured from $(0, 0, \pm\sqrt{\frac{A^4}{\omega^2}+m^2})$.
Using the two-band model, we have also plotted the band structure along the $q_y$ and $q_z$ directions around  the WP at ${\bm k}_{w,z}^{+}$ for $A=0.7$ \AA$^{-1}$, as shown in Fig.~\ref{Fig.5}(c-d). Regardless of the magnitude of $A$, the band structure along $q_z$ direction always remains linear, while it along the $q_y$ directions is effected by non-linear terms  when $A$ is small. 
However, we have checked that this effect diminishes as $A$ increases. 
Consequently, when $A$ is large, the photo-induced WPs become  an ideal linear WP.

The $Z_2$ NL, as a distinctive topological structure within condensed matter systems, offers an appealing platform to explore novel physical properties. 
In the following, based on the effective Hamiltonian derived in this section  and Boltzmann transport theory, we will explore the  anomalous Hall effect and planar Hall effect for the $Z_2$ NL driven by the CPL in different directions. Through this comprehensive analysis, we aim to deepen our understanding of topological phase transitions for $Z_2$ NL under light field manipulation and their potential applications in quantum materials. 

\section{Anomalous Hall effect}\label{sec III}

\begin{figure}
\includegraphics[width=1\columnwidth]{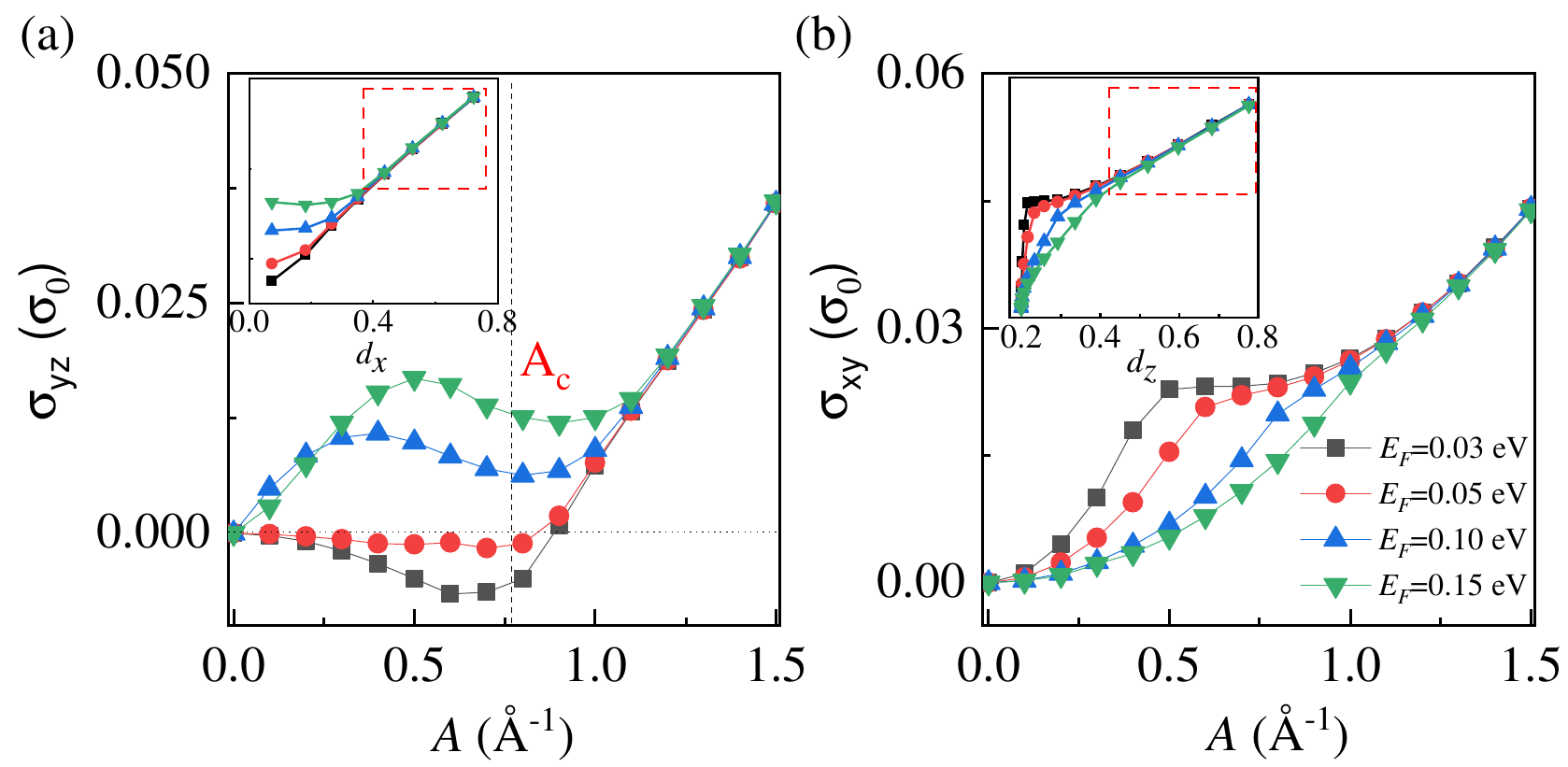}
\caption
{The dependence of AHC on the amplitude for the system driven by the CPL in the (a)$x$ and (b)$z$ direction. The common parameters in use are $m=0.1$ eV, $B=0.5$ T, $\omega=6$ eV, and we have defined the shorthand notation $\sigma_0=\frac{e^2}{h}$. The variation of AHC with the distance between the two WPs is shown in the inset of the figures.}
\label{Fig.3}
\end{figure}

Under the driving of CPL, one of the significant consequences of topological transition is the emergence of the anomalous Hall effect characterized by a non-zero AHC. The CPL breaks the $\mathcal{T}$ symmetry and leads to a photo-induced anomalous Hall effect, which can be adjusted by amplitude and frequency of CPL~\cite{YanZhongbo,DHXu,Oka}. Using linear response theory, the AHC can be obtained as~\cite{Oka}
\begin{eqnarray}
\sigma_{\mu\nu}^{\mathrm{AHE}}=e^2\int\frac{d\bm{k}}{(2\pi)^3}\sum_{n}f_{n}^0(k)\Omega_k^{\rho},
\label{Eq.11}
\end{eqnarray}
where $\mu$, $\nu$, $\rho$=$x$, $y$, $z$, and $n$ is the band index. $\Omega_k^{\rho}$ is the Berry curvature and $f^{0}_{n}(k)$ is the occupation function. The AHC depends not only on the Berry curvature but also on the fermion occupation. In Fig.~\ref{Fig.3}, we show the variation of AHC with amplitudes of light when the CPL is along $x$ and $z$ directions.

The topological phase transitions induced by CPL along different directions result in distinct characteristics of AHC. When the CPL is along $x$ direction, the dependence of $\sigma_{yz}^{\mathrm{AHE}}$ on amplitude $A$ for different Fermi levels are shown in Fig.~\ref{Fig.3}(a). 
As aforementioned, the system driven by the CPL in $x$ direction undergoes two topological phase transitions as amplitude increases. Correspondingly, the characteristics of the AHE are completely different in the two stages. 
Initially, the $Z_2$ NL transforms into a vortex  NL with $A<A_c$, in such case the system features a  torus-shape  Fermi surface when the Fermi energy is small.
However, when the Fermi energy is large, the Fermi surface is ellipsoid. 
Therefore, for $A<A_c$, the variation of $\sigma_{yz}^{\mathrm{AHE}}$ at lower Fermi energy are different from that at higher Fermi energy [see  Fig.~\ref{Fig.3}(a)].

In contrast, when the amplitude of light becomes large, the AHC feature a universal behavior regardless of the Fermi energy, as shown in Fig.~\ref{Fig.3}(a).
For  $A>A_c$, the driven system undergoes the second topological phase transition, transforming into a pair of WPs. The distance between the two WPs is $d_x=2\sqrt{\frac{A^4}{\omega^2}-m^2}$. When $A$ is relatively large, as indicated by the dashed box in the inset in~Fig.~\ref{Fig.3}(a), the distance between the two WPs increases with the increases of $A$, which differs from the WPs generated by driving a normal NL with a large $A$. ~\cite{YanZhongbo,DHXu}. 

For a Weyl semimetal with two well separated WPs, the AHC is  proportional to the distance between the two WPs~\cite{BurkovAA}.  
Therefore, as illustrated in the inset of Fig.~\ref{Fig.3}(a), when $A>A_c$, the $\sigma_{yz}^{\mathrm{AHE}}$ indeed increases linearly with the distance $d_x$ when $A$ is large enough or the Fermi energy is small. 
Furthermore, since the distance $d_x$ is mainly controlled by the amplitude $A$ and increases from zeros, it can be expanded as $d_x\propto(A-A_c)^{1/2}$ around the critical point $A_c$. This means that for $E_F$=0 and $T$=0, the AHC $\sigma_{yz}^{\mathrm{AHE}}$ is approximately proportional to the $(A-A_c)^{1/2}$  around the critical point $A_c$. However, for non-zero Fermi levels, a larger amplitude is required to transform  the  Fermi surface into a pair of separated spheres. Therefore, for a larger Fermi energy, a larger $A$ is needed to observe the proportional relationship between AHE and $d_x$  [see the inset  in Fig.~\ref{Fig.3}(a)].

Then, we discuss the AHC of the $Z_2$ NL  under the driving of CPL in $z$ direction, in such case, the system directly transitions into Weyl semimetal with a pair of WPs at the $k_z$ axial. The distance between the two WPs is $d_z=2\sqrt{\frac{A^4}{\omega^2}+m^2}$, which  also increases from $|m|$ with increasing of $A$.
However when $A$ is small,  the AHC is  not proportional to $d_z$ even for the case of $E_F$=0 and $T$=0, as AHC of the original $Z_2$ NL ($A=0$) is zero.
Therefore, the AHC show a nonanalytical behavior for small $A$, as shown in  Fig.~\ref{Fig.3}(b).
When $A$ is large enough, a linear relationship between $\sigma_{xy}^{\mathrm{AHE}}$ and $d_z$ appears [see Fig.~\ref{Fig.3}(b)]. 
Further, regardless of the direction of the incident  light,  the  AHC of the Floquet $Z_2$ NL is always  proportional to the distance between the two WPs when $A$ is large.

\begin{figure*}
\includegraphics[width=1.5\columnwidth]{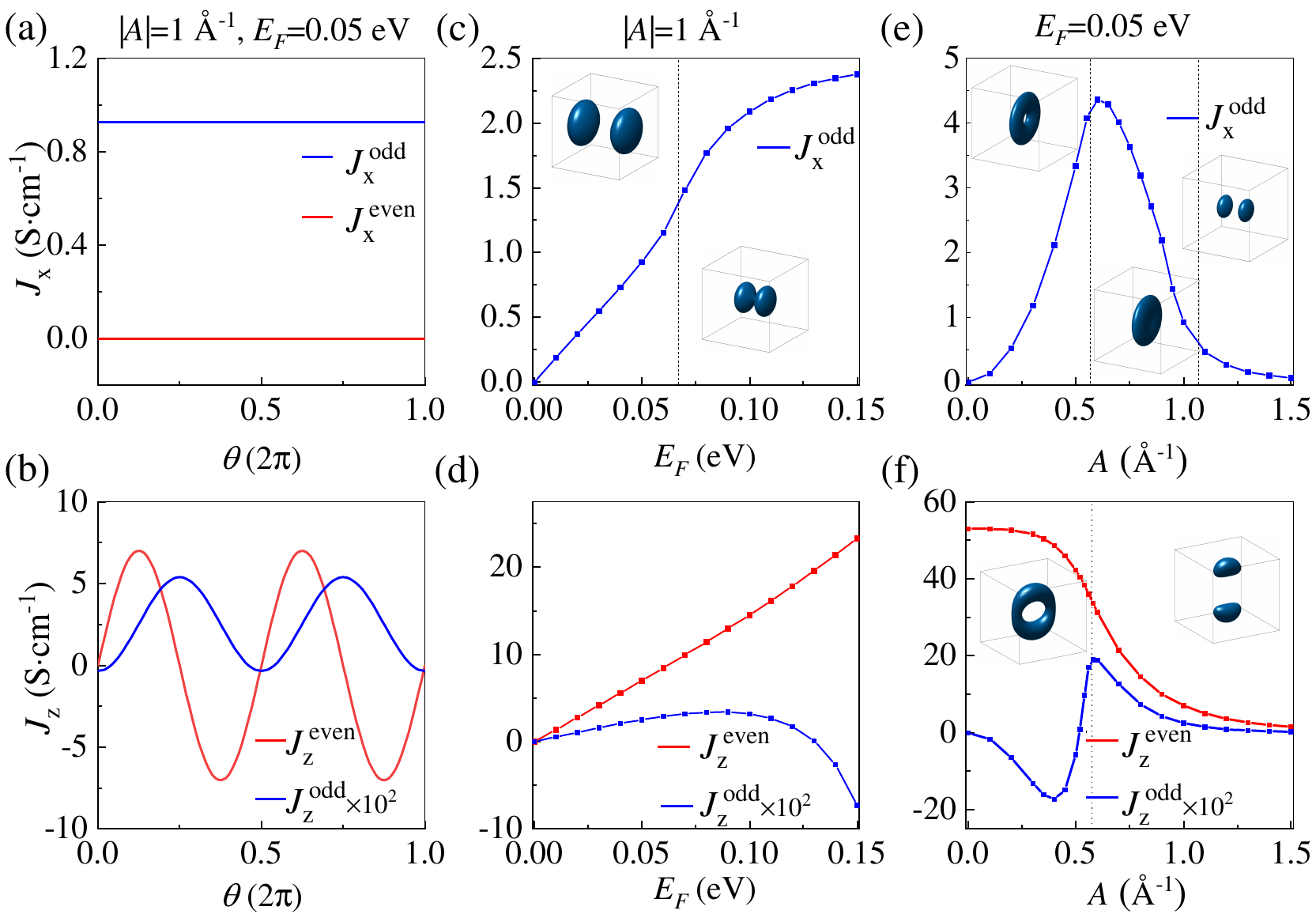}
\caption
{The PHC versus (a-b) $\theta$, (c-d) Fermi level $E_F$ and (e-f) the amplitudes of CPL.  The incident light is along the (a) (c) (e) $x$ direction and (b)(d)(f) $z$ direction, respectively.  The common parameters in use are $m=0.1$ eV, $B=0.5$ T and $\omega=6$ eV.}
\label{Fig.4}
\end{figure*}

\section{Planer Hall effect}\label{sec IV}
We then study the planar Hall effect for the Z$_2$ NL driven by the CPL based on the semiclassical Boltzmann transport theory~\cite{LLeiSciBull,LiLei}. Keeping only the linear order of electric field $\boldsymbol{E}$ and magnetic field $\boldsymbol{B}$, the PHC based on the parity of time can be  divided into two parts~\cite{LiLei}, 
\begin{eqnarray}
\boldsymbol{J}=\boldsymbol{J}^{\mathrm{odd}}+\boldsymbol{J}^{\mathrm{even}},
\end{eqnarray}
\begin{eqnarray*}
\begin{aligned}
&J_{\mu}^{\mathrm{even}}=e^{3}\tau^{2}\int[d\boldsymbol{k}]v_{\mu}\left[\left(\boldsymbol{v}\times\boldsymbol{B}\right)\cdot\partial_{\boldsymbol{k}}\left(\boldsymbol{v\cdot E}\right)\right]\partial_{\varepsilon}f_{k}^{0},\\
&J_{\mu}^{\mathrm{odd}}=-e^{3}\tau\int[d\boldsymbol{k}]v_{\mu}\left(\boldsymbol{v\cdot E}\right)\left(\boldsymbol{B\cdot\Omega_{k}}\right)\partial_{\varepsilon}f_{k}^{0}\\
&-\frac{e^{2}}{\hbar}\tau\int[d\boldsymbol{k}]\left[v_{\mu}\partial_{\boldsymbol{k}}\left(\boldsymbol{m\cdot B}\right)\cdot\boldsymbol{E}+\partial_{k_{\mu}}\left(\boldsymbol{m\cdot B}\right)\left(\boldsymbol{v\cdot E}\right)\right]\partial_{\varepsilon}f_{k}^{0}\\
&+e^{3}\tau\int[d\boldsymbol{k}]\left[B_{\mu}\boldsymbol{\left(v\cdot E\right)}+v_{\mu}\boldsymbol{\left(B\cdot E\right)}\right]\left(\boldsymbol{v\cdot\Omega_{k}}\right)\partial_{\varepsilon}f_{k}^{0}
\label{Eq.12}
\end{aligned}
\end{eqnarray*}
where $\mu=x,y,z$ is the direction, $v_{\mu}=\partial\epsilon(\bm{k})/\hbar$ is the velocity of electrons, and \textbf{\emph{m}} is the orbit magnetic moment. 
For specifically, in subsequent calculations, we focus on the case that  $\boldsymbol{E} \parallel \boldsymbol{B}$, and the both fields are  perpendicular to the direction of light. 
Due to anisotropic band structure, the PHC of  the Floquet $Z_2$ NL also has strongly dependence on the direction of the incident light.

%We show the PHC for $Z_2$ NL driven by the CPL in Fig.~\ref{Fig.4}. 

We first investigated the PHC of the system with CPL along $x$ direction. In this case, $\boldsymbol{E}$ and 
$\boldsymbol{B}$ are  confined in the $y$-$z$ plane, i.e. $\boldsymbol{B}=B(0,\cos\theta,\sin\theta)$ and $\boldsymbol{E}=E(0,\cos\theta,\sin\theta)$. 
From Hamiltonian (\ref{Eq.5}), one observes that while the  $\mathcal{T}$ symmetry is broken by CPL, the system still keeps $\mathcal{C}_{2y}\mathcal{T}=\tau_0\sigma_z\mathcal{K}$ symmetry.
This means that the PHC $J_x$ would only have the term of $J_x^{\mathrm{odd}}$, as $J_x^{\mathrm{even}}$ is forbidden by  $\mathcal{C}_{2y}\mathcal{T}$.
In Fig.~\ref{Fig.4}(a), we plot the variation of $J_x$ with  $\theta$, when the amplitude is $A$=1 $\mathrm{\AA^{-1}}$ and Fermi level is $E_F$=0.05 eV, where the $Z_2$ NL has already transitioned into a pair of WPs, and the corresponding Fermi surface consists of two spheres.
From Fig.~\ref{Fig.4}(a), we indeed have $J_x^{\mathrm{even}}=0$ regardless of the value of $\theta$, and $J_x^{\mathrm{odd}}$ is finite. 
Interestingly, $J_x^{\mathrm{odd}}$  also does not depend on  $\theta$, as the effective Hamiltonian  (\ref{Eq.5}) is isotropic in the $k_y$-$k_z$ planes.

The dependence  of the PHC on Fermi level $E_F$ with fixed $A$ ($A$=1 $\mathrm{\AA^{-1}}$) and $\theta$ ($\theta=0^ {\circ}$) is shown in Fig.~\ref{Fig.4}(c). 
When $E_F$ is small, the Fermi surface consists of two separate spheres, and  $J_x^{\mathrm{odd}}$  approximately linearly increases with $E_F$.
At a critical point $E_F^c$, a Lifshitz phase transition happens and the two separated Fermi spheres merge into a  single one [see Fig.~\ref{Fig.4}(c)].
For $E_F>E_F^c$, while  the PHC  $J_x^{\mathrm{odd}}$ still increases with  $E_F$, but gradually deflects from linear increase. 
In contrast, for $E_F=0.05$ eV, when the amplitude increases, the Fermi surface of the  system evolves from a torus-shape to a closed surface and eventually transforms into a pair of spheres. 
This makes the variation  of the $J_x^{\mathrm{odd}}$ on $A$  not a monotone increasing function, which increases at first and then decreases, as shown in Fig.~\ref{Fig.4}(e).

Then, we consider the case that the  CPL is along the $z$ direction, and the $\boldsymbol{E}$ and $\boldsymbol{B}$ are  confined in the $x$-$y$ plane, i.e.  $\boldsymbol{E}=E(\cos\theta,\sin\theta,0)$ and $\boldsymbol{B}=B(\cos\theta,\sin\theta,0)$.
In  Fig.~\ref{Fig.4}(b), we plot the $J_z$ as a function of  $\theta$. We find that both $J_z^{\mathrm{even}}$ and $J_z^{\mathrm{odd}}$ are finite, and have a period of $\pi$ owing to the setup of $\boldsymbol{E} || \boldsymbol{B}$.
%, owing to terms of $\sin\theta\cos\theta$ and $\cos^2\theta$-$\sin^2\theta$ induced by the $\boldsymbol{B}$ and $\boldsymbol{E}$~\cite{Nandy,MaDa}. 
For $\theta=\pi/4$, we calculate the PHC as a function of $E_F$ and $A$. 
When the Fermi level is small, both $J_z^{\mathrm{even}}$ and $J_z^{\mathrm{odd}}$ almost linearly increase with $E_F$ [see Fig.~\ref{Fig.4}(d)].  However, when $E_F$ is large,  an opposite  trend in the variation of $J_z^{\mathrm{even}}$ and $J_z^{\mathrm{odd}}$ is observed.
For $E_F=0.05$ eV, as $A$ increases, we find that $J_z^{\mathrm{even}}$ monotonically decreases, while $J_z^{\mathrm{odd}}$ feature a complicated behavior, as shown in  Fig.~\ref{Fig.4}(f).

\section{Summary\label{sec:V}}
In this work, based on the $k\cdot p$ model and Floquet theory, we study the topological phase transition, the anomalous and planar Hall effects of the  $Z_2$ NL driven by the CPL in different directions. When the incident  CPL is along $x$ direction, the system undergoes two topological phase transitions: from the  $Z_2$ NL to vortex NL and then to a pair of WPs. This provides a convenient method for obtaining  vortex  NL. However, when the CPL is along the $y$ or $z$ direction, the system will directly transition into a pair of WPs located on the $k_y$ or $k_z$ axis. Moreover, as the amplitude of light increases, the distance between the two photo-induced WPs continuously increases, which makes the AHC of the driven $Z_2$ NL increases with the amplitude of light. Besides, the PHC of the Floquet $Z_2$ NL can  also be  modulated by the direction and amplitude of the incident light. 
These findings deepen our understanding of the Floquet engineering on the $Z_2$ NL, offer different methods to realize  vortex  NL, and may be useful for designing novel AHC and PHC devices.

%When the magnetic and electric field are parallel, and perpendicular to the direction of CPL, and CPL is along $x$ direction, the $J_x^{\mathrm{odd}}$ is zero and $J_x^{\mathrm{even}}$ does not vary with angle of magnetic field due to the $\mathcal{C}_{2y}\mathcal{T}$ symmetry of the driven system. However, for the system driven by CPL in $z$ direction, both $J_z^{\mathrm{odd}}$ and $J_z^{\mathrm{even}}$ have a period of $\pi$ by varying the direction of $\boldsymbol{B}$. Further, regardless of the direction of incident light, when $A$ is large, the dispersion relation of the photo-induced WP is primarily determined by the linear term and the WP behaves like an ideal WP. As $A$ increases, the symmetry of driven system increases, leading to a gradual decrease of PHC towards zeros.

\begin{acknowledgments}
This work was supported by the National Natural Science Foundation
of China (Grant Nos. 12474040, 12474151, and 12347101), the Natural Science Foundation of Chongqing (Grant No. CSTB2022NSCQ-MSX0568) and Beijing National Laboratory for Condensed Matter Physics (No. 2024BNLCMPKF025).
\end{acknowledgments}

\bibliography{ref2}

\begin{thebibliography}{87}%
\makeatletter
\providecommand \@ifxundefined [1]{%
 \@ifx{#1\undefined}
}%
\providecommand \@ifnum [1]{%
 \ifnum #1\expandafter \@firstoftwo
 \else \expandafter \@secondoftwo
 \fi
}%
\providecommand \@ifx [1]{%
 \ifx #1\expandafter \@firstoftwo
 \else \expandafter \@secondoftwo
 \fi
}%
\providecommand \natexlab [1]{#1}%
\providecommand \enquote  [1]{``#1''}%
\providecommand \bibnamefont  [1]{#1}%
\providecommand \bibfnamefont [1]{#1}%
\providecommand \citenamefont [1]{#1}%
\providecommand \href@noop [0]{\@secondoftwo}%
\providecommand \href [0]{\begingroup \@sanitize@url \@href}%
\providecommand \@href[1]{\@@startlink{#1}\@@href}%
\providecommand \@@href[1]{\endgroup#1\@@endlink}%
\providecommand \@sanitize@url [0]{\catcode `\\12\catcode `\$12\catcode
  `\&12\catcode `\#12\catcode `\^12\catcode `\_12\catcode `\%12\relax}%
\providecommand \@@startlink[1]{}%
\providecommand \@@endlink[0]{}%
\providecommand \url  [0]{\begingroup\@sanitize@url \@url }%
\providecommand \@url [1]{\endgroup\@href {#1}{\urlprefix }}%
\providecommand \urlprefix  [0]{URL }%
\providecommand \Eprint [0]{\href }%
\providecommand \doibase [0]{https://doi.org/}%
\providecommand \selectlanguage [0]{\@gobble}%
\providecommand \bibinfo  [0]{\@secondoftwo}%
\providecommand \bibfield  [0]{\@secondoftwo}%
\providecommand \translation [1]{[#1]}%
\providecommand \BibitemOpen [0]{}%
\providecommand \bibitemStop [0]{}%
\providecommand \bibitemNoStop [0]{.\EOS\space}%
\providecommand \EOS [0]{\spacefactor3000\relax}%
\providecommand \BibitemShut  [1]{\csname bibitem#1\endcsname}%
\let\auto@bib@innerbib\@empty
%</preamble>
\bibitem [{\citenamefont {Burkov}(2016)}]{Burkov1}%
  \BibitemOpen
  \bibfield  {author} {\bibinfo {author} {\bibfnamefont {A.~A.}\ \bibnamefont
  {Burkov}},\ }\bibfield  {title} {\bibinfo {title} {{Topological
  semimetals}},\ }\href {https://doi.org/10.1038/nmat4788} {\bibfield
  {journal} {\bibinfo  {journal} {Nat. Mater.}\ }\textbf {\bibinfo {volume}
  {15}},\ \bibinfo {pages} {1145} (\bibinfo {year} {2016})}\BibitemShut
  {NoStop}%
\bibitem [{\citenamefont {Armitage}\ \emph {et~al.}(2018)\citenamefont
  {Armitage}, \citenamefont {Mele},\ and\ \citenamefont {Vishwanath}}]{RMPSM}%
  \BibitemOpen
  \bibfield  {author} {\bibinfo {author} {\bibfnamefont {N.}~\bibnamefont
  {Armitage}}, \bibinfo {author} {\bibfnamefont {E.}~\bibnamefont {Mele}},\
  and\ \bibinfo {author} {\bibfnamefont {A.}~\bibnamefont {Vishwanath}},\
  }\bibfield  {title} {\bibinfo {title} {{Weyl and Dirac semimetals in
  three-dimensional solids}},\ }\href
  {https://doi.org/10.1103/RevModPhys.90.015001} {\bibfield  {journal}
  {\bibinfo  {journal} {Rev. Mod. Phys.}\ }\textbf {\bibinfo {volume} {90}},\
  \bibinfo {pages} {015001} (\bibinfo {year} {2018})}\BibitemShut {NoStop}%
\bibitem [{\citenamefont {Yu}\ \emph {et~al.}(2022)\citenamefont {Yu},
  \citenamefont {Zhang}, \citenamefont {Liu}, \citenamefont {Wu}, \citenamefont
  {Li}, \citenamefont {Zhang}, \citenamefont {Yang},\ and\ \citenamefont
  {Yao}}]{YuSciBull}%
  \BibitemOpen
  \bibfield  {author} {\bibinfo {author} {\bibfnamefont {Z.-M.}\ \bibnamefont
  {Yu}}, \bibinfo {author} {\bibfnamefont {Z.}~\bibnamefont {Zhang}}, \bibinfo
  {author} {\bibfnamefont {G.-B.}\ \bibnamefont {Liu}}, \bibinfo {author}
  {\bibfnamefont {W.}~\bibnamefont {Wu}}, \bibinfo {author} {\bibfnamefont
  {X.-P.}\ \bibnamefont {Li}}, \bibinfo {author} {\bibfnamefont {R.-W.}\
  \bibnamefont {Zhang}}, \bibinfo {author} {\bibfnamefont {S.~A.}\ \bibnamefont
  {Yang}},\ and\ \bibinfo {author} {\bibfnamefont {Y.}~\bibnamefont {Yao}},\
  }\bibfield  {title} {\bibinfo {title} {{Encyclopedia of emergent particles in
  three-dimensional crystals}},\ }\href
  {https://doi.org/10.1016/j.scib.2021.10.023} {\bibfield  {journal} {\bibinfo
  {journal} {Sci. Bull.}\ }\textbf {\bibinfo {volume} {64}},\ \bibinfo {pages}
  {375} (\bibinfo {year} {2022})}\BibitemShut {NoStop}%
\bibitem [{\citenamefont {Wang}\ \emph {et~al.}(2012)\citenamefont {Wang},
  \citenamefont {Sun}, \citenamefont {Chen}, \citenamefont {Franchini},
  \citenamefont {Xu}, \citenamefont {Weng}, \citenamefont {Dai},\ and\
  \citenamefont {Fang}}]{WangZhijun}%
  \BibitemOpen
  \bibfield  {author} {\bibinfo {author} {\bibfnamefont {Z.}~\bibnamefont
  {Wang}}, \bibinfo {author} {\bibfnamefont {Y.}~\bibnamefont {Sun}}, \bibinfo
  {author} {\bibfnamefont {X.-Q.}\ \bibnamefont {Chen}}, \bibinfo {author}
  {\bibfnamefont {C.}~\bibnamefont {Franchini}}, \bibinfo {author}
  {\bibfnamefont {G.}~\bibnamefont {Xu}}, \bibinfo {author} {\bibfnamefont
  {H.}~\bibnamefont {Weng}}, \bibinfo {author} {\bibfnamefont {X.}~\bibnamefont
  {Dai}},\ and\ \bibinfo {author} {\bibfnamefont {Z.}~\bibnamefont {Fang}},\
  }\bibfield  {title} {\bibinfo {title} {{Dirac semimetal and topological phase
  transitions in ${A}_{3}$Bi ($A=\text{Na}$, K, Rb)}},\ }\href
  {https://doi.org/10.1103/PhysRevB.85.195320} {\bibfield  {journal} {\bibinfo
  {journal} {Phys. Rev. B}\ }\textbf {\bibinfo {volume} {85}},\ \bibinfo
  {pages} {195320} (\bibinfo {year} {2012})}\BibitemShut {NoStop}%
\bibitem [{\citenamefont {Wang}\ \emph
  {et~al.}(2013{\natexlab{a}})\citenamefont {Wang}, \citenamefont {Weng},
  \citenamefont {Wu}, \citenamefont {Dai},\ and\ \citenamefont
  {Fang}}]{WangZhijun2}%
  \BibitemOpen
  \bibfield  {author} {\bibinfo {author} {\bibfnamefont {Z.}~\bibnamefont
  {Wang}}, \bibinfo {author} {\bibfnamefont {H.}~\bibnamefont {Weng}}, \bibinfo
  {author} {\bibfnamefont {Q.}~\bibnamefont {Wu}}, \bibinfo {author}
  {\bibfnamefont {X.}~\bibnamefont {Dai}},\ and\ \bibinfo {author}
  {\bibfnamefont {Z.}~\bibnamefont {Fang}},\ }\bibfield  {title} {\bibinfo
  {title} {{Three-dimensional Dirac semimetal and quantum transport in
  Cd${}_{3}$As${}_{2}$}},\ }\href {https://doi.org/10.1103/PhysRevB.88.125427}
  {\bibfield  {journal} {\bibinfo  {journal} {Phys. Rev. B}\ }\textbf {\bibinfo
  {volume} {88}},\ \bibinfo {pages} {125427} (\bibinfo {year}
  {2013}{\natexlab{a}})}\BibitemShut {NoStop}%
\bibitem [{\citenamefont {Neupane}\ \emph {et~al.}(2014)\citenamefont
  {Neupane}, \citenamefont {Xu}, \citenamefont {Sankar}, \citenamefont
  {Alidoust}, \citenamefont {Bian}, \citenamefont {Liu}, \citenamefont
  {Belopolski}, \citenamefont {Chang}, \citenamefont {Jeng}, \citenamefont
  {Lin}, \citenamefont {Bansil}, \citenamefont {Chou},\ and\ \citenamefont
  {Hasan}}]{Madhab}%
  \BibitemOpen
  \bibfield  {author} {\bibinfo {author} {\bibfnamefont {M.}~\bibnamefont
  {Neupane}}, \bibinfo {author} {\bibfnamefont {S.-Y.}\ \bibnamefont {Xu}},
  \bibinfo {author} {\bibfnamefont {R.}~\bibnamefont {Sankar}}, \bibinfo
  {author} {\bibfnamefont {N.}~\bibnamefont {Alidoust}}, \bibinfo {author}
  {\bibfnamefont {G.}~\bibnamefont {Bian}}, \bibinfo {author} {\bibfnamefont
  {C.}~\bibnamefont {Liu}}, \bibinfo {author} {\bibfnamefont {I.}~\bibnamefont
  {Belopolski}}, \bibinfo {author} {\bibfnamefont {T.-R.}\ \bibnamefont
  {Chang}}, \bibinfo {author} {\bibfnamefont {H.-T.}\ \bibnamefont {Jeng}},
  \bibinfo {author} {\bibfnamefont {H.}~\bibnamefont {Lin}}, \bibinfo {author}
  {\bibfnamefont {A.}~\bibnamefont {Bansil}}, \bibinfo {author} {\bibfnamefont
  {F.}~\bibnamefont {Chou}},\ and\ \bibinfo {author} {\bibfnamefont {M.~Z.}\
  \bibnamefont {Hasan}},\ }\bibfield  {title} {\bibinfo {title} {{Observation
  of a three-dimensional topological Dirac semimetal phase in high-mobility
  Cd${}_{3}$As${}_{2}$}},\ }\href {https://doi.org/10.1038/ncomms4786}
  {\bibfield  {journal} {\bibinfo  {journal} {Nat. Commun.}\ }\textbf {\bibinfo
  {volume} {5}},\ \bibinfo {pages} {3786} (\bibinfo {year} {2014})}\BibitemShut
  {NoStop}%
\bibitem [{\citenamefont {Liu}\ \emph {et~al.}(2014)\citenamefont {Liu},
  \citenamefont {Jiang}, \citenamefont {Zhou}, \citenamefont {Wang},
  \citenamefont {Zhang}, \citenamefont {Weng}, \citenamefont {Prabhakaran},
  \citenamefont {Mo}, \citenamefont {Peng}, \citenamefont {Dudin},
  \citenamefont {Kim}, \citenamefont {Hoesch}, \citenamefont {Fang},
  \citenamefont {Dai}, \citenamefont {Shen}, \citenamefont {D.~L.~Feng},\ and\
  \citenamefont {Chen}}]{ZKLiu}%
  \BibitemOpen
  \bibfield  {author} {\bibinfo {author} {\bibfnamefont {Z.~K.}\ \bibnamefont
  {Liu}}, \bibinfo {author} {\bibfnamefont {J.}~\bibnamefont {Jiang}}, \bibinfo
  {author} {\bibfnamefont {B.}~\bibnamefont {Zhou}}, \bibinfo {author}
  {\bibfnamefont {Z.~J.}\ \bibnamefont {Wang}}, \bibinfo {author}
  {\bibfnamefont {Y.}~\bibnamefont {Zhang}}, \bibinfo {author} {\bibfnamefont
  {H.~M.}\ \bibnamefont {Weng}}, \bibinfo {author} {\bibfnamefont
  {D.}~\bibnamefont {Prabhakaran}}, \bibinfo {author} {\bibfnamefont {S.-K.}\
  \bibnamefont {Mo}}, \bibinfo {author} {\bibfnamefont {H.}~\bibnamefont
  {Peng}}, \bibinfo {author} {\bibfnamefont {P.}~\bibnamefont {Dudin}},
  \bibinfo {author} {\bibfnamefont {T.}~\bibnamefont {Kim}}, \bibinfo {author}
  {\bibfnamefont {M.}~\bibnamefont {Hoesch}}, \bibinfo {author} {\bibfnamefont
  {Z.}~\bibnamefont {Fang}}, \bibinfo {author} {\bibfnamefont {X.}~\bibnamefont
  {Dai}}, \bibinfo {author} {\bibfnamefont {Z.~X.}\ \bibnamefont {Shen}},
  \bibinfo {author} {\bibfnamefont {Z.~H.}\ \bibnamefont {D.~L.~Feng}},\ and\
  \bibinfo {author} {\bibfnamefont {Y.~L.}\ \bibnamefont {Chen}},\ }\bibfield
  {title} {\bibinfo {title} {{A stable three-dimensional topological Dirac
  semimetal Cd${}_{3}$As${}_{2}$}},\ }\href {https://doi.org/10.1038/nmat3990}
  {\bibfield  {journal} {\bibinfo  {journal} {Nat. Mater.}\ }\textbf {\bibinfo
  {volume} {13}},\ \bibinfo {pages} {577} (\bibinfo {year} {2014})}\BibitemShut
  {NoStop}%
\bibitem [{\citenamefont {Young}\ \emph {et~al.}(2012)\citenamefont {Young},
  \citenamefont {Zaheer}, \citenamefont {Teo}, \citenamefont {Kane},
  \citenamefont {Mele},\ and\ \citenamefont {Rappe}}]{SMYoung}%
  \BibitemOpen
  \bibfield  {author} {\bibinfo {author} {\bibfnamefont {S.~M.}\ \bibnamefont
  {Young}}, \bibinfo {author} {\bibfnamefont {S.}~\bibnamefont {Zaheer}},
  \bibinfo {author} {\bibfnamefont {J.~C.~Y.}\ \bibnamefont {Teo}}, \bibinfo
  {author} {\bibfnamefont {C.~L.}\ \bibnamefont {Kane}}, \bibinfo {author}
  {\bibfnamefont {E.~J.}\ \bibnamefont {Mele}},\ and\ \bibinfo {author}
  {\bibfnamefont {A.~M.}\ \bibnamefont {Rappe}},\ }\bibfield  {title} {\bibinfo
  {title} {{Dirac Semimetal in Three Dimensions}},\ }\href
  {https://doi.org/10.1103/PhysRevLett.108.140405} {\bibfield  {journal}
  {\bibinfo  {journal} {Phys. Rev. Lett.}\ }\textbf {\bibinfo {volume} {108}},\
  \bibinfo {pages} {140405} (\bibinfo {year} {2012})}\BibitemShut {NoStop}%
\bibitem [{\citenamefont {Wan}\ \emph {et~al.}(2011)\citenamefont {Wan},
  \citenamefont {Turner}, \citenamefont {Vishwanath},\ and\ \citenamefont
  {Savrasov}}]{WanXiangang}%
  \BibitemOpen
  \bibfield  {author} {\bibinfo {author} {\bibfnamefont {X.}~\bibnamefont
  {Wan}}, \bibinfo {author} {\bibfnamefont {A.~M.}\ \bibnamefont {Turner}},
  \bibinfo {author} {\bibfnamefont {A.}~\bibnamefont {Vishwanath}},\ and\
  \bibinfo {author} {\bibfnamefont {S.~Y.}\ \bibnamefont {Savrasov}},\
  }\bibfield  {title} {\bibinfo {title} {{Topological semimetal and Fermi-arc
  surface states in the electronic structure of pyrochlore iridates}},\ }\href
  {https://doi.org/10.1103/PhysRevB.83.205101} {\bibfield  {journal} {\bibinfo
  {journal} {Phys. Rev. B}\ }\textbf {\bibinfo {volume} {83}},\ \bibinfo
  {pages} {205101} (\bibinfo {year} {2011})}\BibitemShut {NoStop}%
\bibitem [{\citenamefont {Lv}\ \emph {et~al.}(2015{\natexlab{a}})\citenamefont
  {Lv}, \citenamefont {Weng}, \citenamefont {Fu}, \citenamefont {Wang},
  \citenamefont {Miao}, \citenamefont {Ma}, \citenamefont {Richard},
  \citenamefont {Huang}, \citenamefont {Zhao}, \citenamefont {Chen},
  \citenamefont {Fang}, \citenamefont {Dai}, \citenamefont {Qian},\ and\
  \citenamefont {Ding}}]{LvBQ}%
  \BibitemOpen
  \bibfield  {author} {\bibinfo {author} {\bibfnamefont {B.~Q.}\ \bibnamefont
  {Lv}}, \bibinfo {author} {\bibfnamefont {H.~M.}\ \bibnamefont {Weng}},
  \bibinfo {author} {\bibfnamefont {B.~B.}\ \bibnamefont {Fu}}, \bibinfo
  {author} {\bibfnamefont {X.~P.}\ \bibnamefont {Wang}}, \bibinfo {author}
  {\bibfnamefont {H.}~\bibnamefont {Miao}}, \bibinfo {author} {\bibfnamefont
  {J.}~\bibnamefont {Ma}}, \bibinfo {author} {\bibfnamefont {P.}~\bibnamefont
  {Richard}}, \bibinfo {author} {\bibfnamefont {X.~C.}\ \bibnamefont {Huang}},
  \bibinfo {author} {\bibfnamefont {L.~X.}\ \bibnamefont {Zhao}}, \bibinfo
  {author} {\bibfnamefont {G.~F.}\ \bibnamefont {Chen}}, \bibinfo {author}
  {\bibfnamefont {Z.}~\bibnamefont {Fang}}, \bibinfo {author} {\bibfnamefont
  {X.}~\bibnamefont {Dai}}, \bibinfo {author} {\bibfnamefont {T.}~\bibnamefont
  {Qian}},\ and\ \bibinfo {author} {\bibfnamefont {H.}~\bibnamefont {Ding}},\
  }\bibfield  {title} {\bibinfo {title} {{Experimental Discovery of Weyl
  Semimetal TaAs}},\ }\href {https://doi.org/10.1103/PhysRevX.5.031013}
  {\bibfield  {journal} {\bibinfo  {journal} {Phys. Rev. X}\ }\textbf {\bibinfo
  {volume} {5}},\ \bibinfo {pages} {031013} (\bibinfo {year}
  {2015}{\natexlab{a}})}\BibitemShut {NoStop}%
\bibitem [{\citenamefont {Lv}\ \emph {et~al.}(2015{\natexlab{b}})\citenamefont
  {Lv}, \citenamefont {Xu}, \citenamefont {Weng}, \citenamefont {Ma},
  \citenamefont {Richard}, \citenamefont {Huang}, \citenamefont {Zhao},
  \citenamefont {Chen}, \citenamefont {Matt}, \citenamefont {Bisti},
  \citenamefont {Strocov}, \citenamefont {Mesot}, \citenamefont {Fang},
  \citenamefont {Dai}, \citenamefont {Qian}, \citenamefont {Shi},\ and\
  \citenamefont {Ding}}]{BQLv2}%
  \BibitemOpen
  \bibfield  {author} {\bibinfo {author} {\bibfnamefont {B.~Q.}\ \bibnamefont
  {Lv}}, \bibinfo {author} {\bibfnamefont {N.}~\bibnamefont {Xu}}, \bibinfo
  {author} {\bibfnamefont {H.~M.}\ \bibnamefont {Weng}}, \bibinfo {author}
  {\bibfnamefont {J.~Z.}\ \bibnamefont {Ma}}, \bibinfo {author} {\bibfnamefont
  {P.}~\bibnamefont {Richard}}, \bibinfo {author} {\bibfnamefont {X.~C.}\
  \bibnamefont {Huang}}, \bibinfo {author} {\bibfnamefont {L.~X.}\ \bibnamefont
  {Zhao}}, \bibinfo {author} {\bibfnamefont {G.~F.}\ \bibnamefont {Chen}},
  \bibinfo {author} {\bibfnamefont {C.~E.}\ \bibnamefont {Matt}}, \bibinfo
  {author} {\bibfnamefont {F.}~\bibnamefont {Bisti}}, \bibinfo {author}
  {\bibfnamefont {V.~N.}\ \bibnamefont {Strocov}}, \bibinfo {author}
  {\bibfnamefont {J.}~\bibnamefont {Mesot}}, \bibinfo {author} {\bibfnamefont
  {Z.}~\bibnamefont {Fang}}, \bibinfo {author} {\bibfnamefont {X.}~\bibnamefont
  {Dai}}, \bibinfo {author} {\bibfnamefont {T.}~\bibnamefont {Qian}}, \bibinfo
  {author} {\bibfnamefont {M.}~\bibnamefont {Shi}},\ and\ \bibinfo {author}
  {\bibfnamefont {H.}~\bibnamefont {Ding}},\ }\bibfield  {title} {\bibinfo
  {title} {{Observation of Weyl nodes in TaAs}},\ }\href
  {https://doi.org/10.1038/nphys3426} {\bibfield  {journal} {\bibinfo
  {journal} {Nat. Phys.}\ }\textbf {\bibinfo {volume} {11}},\ \bibinfo {pages}
  {724} (\bibinfo {year} {2015}{\natexlab{b}})}\BibitemShut {NoStop}%
\bibitem [{\citenamefont {Burkov}\ and\ \citenamefont
  {Balents}(2011)}]{BurkovAA}%
  \BibitemOpen
  \bibfield  {author} {\bibinfo {author} {\bibfnamefont {A.~A.}\ \bibnamefont
  {Burkov}}\ and\ \bibinfo {author} {\bibfnamefont {L.}~\bibnamefont
  {Balents}},\ }\bibfield  {title} {\bibinfo {title} {{Weyl Semimetal in a
  Topological Insulator Multilayer}},\ }\href
  {https://doi.org/10.1103/PhysRevLett.107.127205} {\bibfield  {journal}
  {\bibinfo  {journal} {Phys. Rev. Lett.}\ }\textbf {\bibinfo {volume} {107}},\
  \bibinfo {pages} {127205} (\bibinfo {year} {2011})}\BibitemShut {NoStop}%
\bibitem [{\citenamefont {Yang}\ \emph {et~al.}(2011)\citenamefont {Yang},
  \citenamefont {Lu},\ and\ \citenamefont {Ran}}]{YangKaiYu}%
  \BibitemOpen
  \bibfield  {author} {\bibinfo {author} {\bibfnamefont {K.-Y.}\ \bibnamefont
  {Yang}}, \bibinfo {author} {\bibfnamefont {Y.-M.}\ \bibnamefont {Lu}},\ and\
  \bibinfo {author} {\bibfnamefont {Y.}~\bibnamefont {Ran}},\ }\bibfield
  {title} {\bibinfo {title} {{Quantum Hall effects in a Weyl semimetal:
  Possible application in pyrochlore iridates}},\ }\href
  {https://doi.org/10.1103/PhysRevB.84.075129} {\bibfield  {journal} {\bibinfo
  {journal} {Phys. Rev. B}\ }\textbf {\bibinfo {volume} {84}},\ \bibinfo
  {pages} {075129} (\bibinfo {year} {2011})}\BibitemShut {NoStop}%
\bibitem [{\citenamefont {Weng}\ \emph
  {et~al.}(2015{\natexlab{a}})\citenamefont {Weng}, \citenamefont {Fang},
  \citenamefont {Fang}, \citenamefont {Bernevig},\ and\ \citenamefont
  {Dai}}]{WengHongming2}%
  \BibitemOpen
  \bibfield  {author} {\bibinfo {author} {\bibfnamefont {H.}~\bibnamefont
  {Weng}}, \bibinfo {author} {\bibfnamefont {C.}~\bibnamefont {Fang}}, \bibinfo
  {author} {\bibfnamefont {Z.}~\bibnamefont {Fang}}, \bibinfo {author}
  {\bibfnamefont {B.~A.}\ \bibnamefont {Bernevig}},\ and\ \bibinfo {author}
  {\bibfnamefont {X.}~\bibnamefont {Dai}},\ }\bibfield  {title} {\bibinfo
  {title} {{Weyl Semimetal Phase in Noncentrosymmetric Transition-Metal
  Monophosphides}},\ }\href {https://doi.org/10.1103/PhysRevX.5.011029}
  {\bibfield  {journal} {\bibinfo  {journal} {Phys. Rev. X}\ }\textbf {\bibinfo
  {volume} {5}},\ \bibinfo {pages} {011029} (\bibinfo {year}
  {2015}{\natexlab{a}})}\BibitemShut {NoStop}%
\bibitem [{\citenamefont {Chan}\ \emph
  {et~al.}(2016{\natexlab{a}})\citenamefont {Chan}, \citenamefont {Chiu},
  \citenamefont {Chou},\ and\ \citenamefont {Schnyder}}]{ChanYH}%
  \BibitemOpen
  \bibfield  {author} {\bibinfo {author} {\bibfnamefont {Y.-H.}\ \bibnamefont
  {Chan}}, \bibinfo {author} {\bibfnamefont {C.-K.}\ \bibnamefont {Chiu}},
  \bibinfo {author} {\bibfnamefont {M.~Y.}\ \bibnamefont {Chou}},\ and\
  \bibinfo {author} {\bibfnamefont {A.~P.}\ \bibnamefont {Schnyder}},\
  }\bibfield  {title} {\bibinfo {title} {{${\mathrm{Ca}}_{3}{\mathrm{P}}_{2}$
  and other topological semimetals with line nodes and drumhead surface
  states}},\ }\href {https://doi.org/10.1103/PhysRevB.93.205132} {\bibfield
  {journal} {\bibinfo  {journal} {Phys. Rev. B}\ }\textbf {\bibinfo {volume}
  {93}},\ \bibinfo {pages} {205132} (\bibinfo {year}
  {2016}{\natexlab{a}})}\BibitemShut {NoStop}%
\bibitem [{\citenamefont {Fang}\ \emph {et~al.}(2016)\citenamefont {Fang},
  \citenamefont {Lu}, \citenamefont {Liu},\ and\ \citenamefont
  {Fu}}]{ChenFang}%
  \BibitemOpen
  \bibfield  {author} {\bibinfo {author} {\bibfnamefont {C.}~\bibnamefont
  {Fang}}, \bibinfo {author} {\bibfnamefont {L.}~\bibnamefont {Lu}}, \bibinfo
  {author} {\bibfnamefont {J.}~\bibnamefont {Liu}},\ and\ \bibinfo {author}
  {\bibfnamefont {L.}~\bibnamefont {Fu}},\ }\bibfield  {title} {\bibinfo
  {title} {{Topological semimetals with helicoid surface states}},\ }\href
  {https://doi.org/10.1038/nphys3782} {\bibfield  {journal} {\bibinfo
  {journal} {Nat. Phys.}\ }\textbf {\bibinfo {volume} {12}},\ \bibinfo {pages}
  {936} (\bibinfo {year} {2016})}\BibitemShut {NoStop}%
\bibitem [{\citenamefont {Chen}\ \emph {et~al.}(2015)\citenamefont {Chen},
  \citenamefont {Xie}, \citenamefont {Yang}, \citenamefont {Pan}, \citenamefont
  {Zhang}, \citenamefont {Cohen},\ and\ \citenamefont {Zhang}}]{YuanpingChen}%
  \BibitemOpen
  \bibfield  {author} {\bibinfo {author} {\bibfnamefont {Y.}~\bibnamefont
  {Chen}}, \bibinfo {author} {\bibfnamefont {Y.}~\bibnamefont {Xie}}, \bibinfo
  {author} {\bibfnamefont {S.~A.}\ \bibnamefont {Yang}}, \bibinfo {author}
  {\bibfnamefont {H.}~\bibnamefont {Pan}}, \bibinfo {author} {\bibfnamefont
  {F.}~\bibnamefont {Zhang}}, \bibinfo {author} {\bibfnamefont {M.~L.}\
  \bibnamefont {Cohen}},\ and\ \bibinfo {author} {\bibfnamefont
  {S.}~\bibnamefont {Zhang}},\ }\bibfield  {title} {\bibinfo {title}
  {{Nanostructured Carbon Allotropes with Weyl-like Loops and Points}},\ }\href
  {https://doi.org/10.1021/acs.nanolett.5b02978} {\bibfield  {journal}
  {\bibinfo  {journal} {Nano Lett.}\ }\textbf {\bibinfo {volume} {10}},\
  \bibinfo {pages} {6974} (\bibinfo {year} {2015})}\BibitemShut {NoStop}%
\bibitem [{\citenamefont {Bzdušek}\ \emph {et~al.}(2016)\citenamefont
  {Bzdušek}, \citenamefont {Wu}, \citenamefont {Rüegg}, \citenamefont
  {Sigrist},\ and\ \citenamefont {Soluyanov}}]{Soluyanov}%
  \BibitemOpen
  \bibfield  {author} {\bibinfo {author} {\bibfnamefont {T.}~\bibnamefont
  {Bzdušek}}, \bibinfo {author} {\bibfnamefont {Q.}~\bibnamefont {Wu}},
  \bibinfo {author} {\bibfnamefont {A.}~\bibnamefont {Rüegg}}, \bibinfo
  {author} {\bibfnamefont {M.}~\bibnamefont {Sigrist}},\ and\ \bibinfo {author}
  {\bibfnamefont {A.~A.}\ \bibnamefont {Soluyanov}},\ }\bibfield  {title}
  {\bibinfo {title} {{Nodal-chain metals}},\ }\href
  {https://doi.org/10.1038/nature19099} {\bibfield  {journal} {\bibinfo
  {journal} {Nature}\ }\textbf {\bibinfo {volume} {538}},\ \bibinfo {pages}
  {75} (\bibinfo {year} {2016})}\BibitemShut {NoStop}%
\bibitem [{\citenamefont {Burkov}\ \emph {et~al.}(2011)\citenamefont {Burkov},
  \citenamefont {Hook},\ and\ \citenamefont {Balents}}]{Burkov}%
  \BibitemOpen
  \bibfield  {author} {\bibinfo {author} {\bibfnamefont {A.~A.}\ \bibnamefont
  {Burkov}}, \bibinfo {author} {\bibfnamefont {M.~D.}\ \bibnamefont {Hook}},\
  and\ \bibinfo {author} {\bibfnamefont {L.}~\bibnamefont {Balents}},\
  }\bibfield  {title} {\bibinfo {title} {{Topological nodal semimetals}},\
  }\href {https://doi.org/10.1103/PhysRevB.84.235126} {\bibfield  {journal}
  {\bibinfo  {journal} {Phys. Rev. B}\ }\textbf {\bibinfo {volume} {84}},\
  \bibinfo {pages} {235126} (\bibinfo {year} {2011})}\BibitemShut {NoStop}%
\bibitem [{\citenamefont {Weng}\ \emph
  {et~al.}(2015{\natexlab{b}})\citenamefont {Weng}, \citenamefont {Liang},
  \citenamefont {Xu}, \citenamefont {Yu}, \citenamefont {Fang}, \citenamefont
  {Dai},\ and\ \citenamefont {Kawazoe}}]{WengHongming}%
  \BibitemOpen
  \bibfield  {author} {\bibinfo {author} {\bibfnamefont {H.}~\bibnamefont
  {Weng}}, \bibinfo {author} {\bibfnamefont {Y.}~\bibnamefont {Liang}},
  \bibinfo {author} {\bibfnamefont {Q.}~\bibnamefont {Xu}}, \bibinfo {author}
  {\bibfnamefont {R.}~\bibnamefont {Yu}}, \bibinfo {author} {\bibfnamefont
  {Z.}~\bibnamefont {Fang}}, \bibinfo {author} {\bibfnamefont {X.}~\bibnamefont
  {Dai}},\ and\ \bibinfo {author} {\bibfnamefont {Y.}~\bibnamefont {Kawazoe}},\
  }\bibfield  {title} {\bibinfo {title} {{Topological node-line semimetal in
  three-dimensional graphene networks}},\ }\href
  {https://doi.org/10.1103/PhysRevB.92.045108} {\bibfield  {journal} {\bibinfo
  {journal} {Phys. Rev. B}\ }\textbf {\bibinfo {volume} {92}},\ \bibinfo
  {pages} {045108} (\bibinfo {year} {2015}{\natexlab{b}})}\BibitemShut
  {NoStop}%
\bibitem [{\citenamefont {Zhang}\ \emph {et~al.}(2018)\citenamefont {Zhang},
  \citenamefont {Yu}, \citenamefont {Lu}, \citenamefont {Sheng}, \citenamefont
  {Yang},\ and\ \citenamefont {Yang1}}]{XMZhang}%
  \BibitemOpen
  \bibfield  {author} {\bibinfo {author} {\bibfnamefont {X.}~\bibnamefont
  {Zhang}}, \bibinfo {author} {\bibfnamefont {Z.-M.}\ \bibnamefont {Yu}},
  \bibinfo {author} {\bibfnamefont {Y.}~\bibnamefont {Lu}}, \bibinfo {author}
  {\bibfnamefont {X.-L.}\ \bibnamefont {Sheng}}, \bibinfo {author}
  {\bibfnamefont {H.~Y.}\ \bibnamefont {Yang}},\ and\ \bibinfo {author}
  {\bibfnamefont {S.~A.}\ \bibnamefont {Yang1}},\ }\bibfield  {title} {\bibinfo
  {title} {{Hybrid nodal loop metal: Unconventional magnetoresponse and
  material realization}},\ }\href {https://doi.org/10.1103/PhysRevB.97.125143}
  {\bibfield  {journal} {\bibinfo  {journal} {Phys. Rev. B}\ }\textbf {\bibinfo
  {volume} {97}},\ \bibinfo {pages} {125143} (\bibinfo {year}
  {2018})}\BibitemShut {NoStop}%
\bibitem [{\citenamefont {Fang}\ \emph {et~al.}(2015)\citenamefont {Fang},
  \citenamefont {Chen}, \citenamefont {Kee},\ and\ \citenamefont
  {Fu}}]{FangChen2}%
  \BibitemOpen
  \bibfield  {author} {\bibinfo {author} {\bibfnamefont {C.}~\bibnamefont
  {Fang}}, \bibinfo {author} {\bibfnamefont {Y.}~\bibnamefont {Chen}}, \bibinfo
  {author} {\bibfnamefont {H.-Y.}\ \bibnamefont {Kee}},\ and\ \bibinfo {author}
  {\bibfnamefont {L.}~\bibnamefont {Fu}},\ }\bibfield  {title} {\bibinfo
  {title} {{Topological nodal line semimetals with and without spin-orbital
  coupling}},\ }\href {https://doi.org/10.1103/PhysRevB.92.081201} {\bibfield
  {journal} {\bibinfo  {journal} {Phys. Rev. B}\ }\textbf {\bibinfo {volume}
  {92}},\ \bibinfo {pages} {081201(R)} (\bibinfo {year} {2015})}\BibitemShut
  {NoStop}%
\bibitem [{\citenamefont {Chiu}\ and\ \citenamefont {Schnyder}(2014)}]{Chiu}%
  \BibitemOpen
  \bibfield  {author} {\bibinfo {author} {\bibfnamefont {C.-K.}\ \bibnamefont
  {Chiu}}\ and\ \bibinfo {author} {\bibfnamefont {A.~P.}\ \bibnamefont
  {Schnyder}},\ }\bibfield  {title} {\bibinfo {title} {{Classification of
  reflection-symmetry-protected topological semimetals and nodal
  superconductors}},\ }\href {https://doi.org/10.1103/PhysRevB.90.205136}
  {\bibfield  {journal} {\bibinfo  {journal} {Phys. Rev. B}\ }\textbf {\bibinfo
  {volume} {90}},\ \bibinfo {pages} {205136} (\bibinfo {year}
  {2014})}\BibitemShut {NoStop}%
\bibitem [{\citenamefont {Kim}\ \emph {et~al.}(2015)\citenamefont {Kim},
  \citenamefont {Wieder}, \citenamefont {Kane},\ and\ \citenamefont
  {Rappe}}]{Youngkuk}%
  \BibitemOpen
  \bibfield  {author} {\bibinfo {author} {\bibfnamefont {Y.}~\bibnamefont
  {Kim}}, \bibinfo {author} {\bibfnamefont {B.~J.}\ \bibnamefont {Wieder}},
  \bibinfo {author} {\bibfnamefont {C.~L.}\ \bibnamefont {Kane}},\ and\
  \bibinfo {author} {\bibfnamefont {A.~M.}\ \bibnamefont {Rappe}},\ }\bibfield
  {title} {\bibinfo {title} {{Dirac Line Nodes in Inversion-Symmetric
  Crystals}},\ }\href {https://doi.org/10.1103/PhysRevLett.115.036806}
  {\bibfield  {journal} {\bibinfo  {journal} {Phys. Rev. Lett.}\ }\textbf
  {\bibinfo {volume} {115}},\ \bibinfo {pages} {036806} (\bibinfo {year}
  {2015})}\BibitemShut {NoStop}%
\bibitem [{\citenamefont {Zhao}\ \emph {et~al.}(2016)\citenamefont {Zhao},
  \citenamefont {Schnyder},\ and\ \citenamefont {Wang}}]{ZhaoYX}%
  \BibitemOpen
  \bibfield  {author} {\bibinfo {author} {\bibfnamefont {Y.~X.}\ \bibnamefont
  {Zhao}}, \bibinfo {author} {\bibfnamefont {A.~P.}\ \bibnamefont {Schnyder}},\
  and\ \bibinfo {author} {\bibfnamefont {Z.~D.}\ \bibnamefont {Wang}},\
  }\bibfield  {title} {\bibinfo {title} {{Unified Theory of $PT$ and $CP$
  Invariant Topological Metals and Nodal Superconductors}},\ }\href
  {https://doi.org/10.1103/PhysRevLett.116.156402} {\bibfield  {journal}
  {\bibinfo  {journal} {Phys. Rev. Lett.}\ }\textbf {\bibinfo {volume} {116}},\
  \bibinfo {pages} {156402} (\bibinfo {year} {2016})}\BibitemShut {NoStop}%
\bibitem [{\citenamefont {Song}\ \emph {et~al.}(2018)\citenamefont {Song},
  \citenamefont {Zhang},\ and\ \citenamefont {Fang}}]{SongZhida}%
  \BibitemOpen
  \bibfield  {author} {\bibinfo {author} {\bibfnamefont {Z.}~\bibnamefont
  {Song}}, \bibinfo {author} {\bibfnamefont {T.}~\bibnamefont {Zhang}},\ and\
  \bibinfo {author} {\bibfnamefont {C.}~\bibnamefont {Fang}},\ }\bibfield
  {title} {\bibinfo {title} {{Diagnosis for Nonmagnetic Topological Semimetals
  in the Absence of Spin-Orbital Coupling}},\ }\href
  {https://doi.org/10.1103/PhysRevX.8.031069} {\bibfield  {journal} {\bibinfo
  {journal} {Phys. Rev. X}\ }\textbf {\bibinfo {volume} {8}},\ \bibinfo {pages}
  {031069} (\bibinfo {year} {2018})}\BibitemShut {NoStop}%
\bibitem [{\citenamefont {Kawakami}\ \emph {et~al.}(2020)\citenamefont
  {Kawakami}, \citenamefont {Nomura},\ and\ \citenamefont
  {Koshino}}]{Kawakami}%
  \BibitemOpen
  \bibfield  {author} {\bibinfo {author} {\bibfnamefont {T.}~\bibnamefont
  {Kawakami}}, \bibinfo {author} {\bibfnamefont {T.}~\bibnamefont {Nomura}},\
  and\ \bibinfo {author} {\bibfnamefont {M.}~\bibnamefont {Koshino}},\
  }\bibfield  {title} {\bibinfo {title} {{Electronic properties of a
  graphyne-$N$ monolayer and its multilayer: Even-odd effect and topological
  nodal line semimetalic phases}},\ }\href
  {https://doi.org/10.1103/PhysRevB.102.115421} {\bibfield  {journal} {\bibinfo
   {journal} {Phys. Rev. B}\ }\textbf {\bibinfo {volume} {102}},\ \bibinfo
  {pages} {115421} (\bibinfo {year} {2020})}\BibitemShut {NoStop}%
\bibitem [{\citenamefont {Ahn}\ \emph {et~al.}(2018)\citenamefont {Ahn},
  \citenamefont {Kim}, \citenamefont {Kim},\ and\ \citenamefont {Yang}}]{Ahn}%
  \BibitemOpen
  \bibfield  {author} {\bibinfo {author} {\bibfnamefont {J.}~\bibnamefont
  {Ahn}}, \bibinfo {author} {\bibfnamefont {D.}~\bibnamefont {Kim}}, \bibinfo
  {author} {\bibfnamefont {Y.}~\bibnamefont {Kim}},\ and\ \bibinfo {author}
  {\bibfnamefont {B.-J.}\ \bibnamefont {Yang}},\ }\bibfield  {title} {\bibinfo
  {title} {{Band Topology and Linking Structure of Nodal Line Semimetals with
  ${Z}_{2}$ Monopole Charges}},\ }\href
  {https://doi.org/10.1103/PhysRevLett.121.106403} {\bibfield  {journal}
  {\bibinfo  {journal} {Phys. Rev. Lett.}\ }\textbf {\bibinfo {volume} {121}},\
  \bibinfo {pages} {106403} (\bibinfo {year} {2018})}\BibitemShut {NoStop}%
\bibitem [{\citenamefont {Morimoto}\ and\ \citenamefont
  {Furusaki}(2014)}]{Morimoto}%
  \BibitemOpen
  \bibfield  {author} {\bibinfo {author} {\bibfnamefont {T.}~\bibnamefont
  {Morimoto}}\ and\ \bibinfo {author} {\bibfnamefont {A.}~\bibnamefont
  {Furusaki}},\ }\bibfield  {title} {\bibinfo {title} {{Weyl and Dirac
  semimetals with ${\mathbb{Z}}_{2}$ topological charge}},\ }\href
  {https://doi.org/10.1103/PhysRevB.89.235127} {\bibfield  {journal} {\bibinfo
  {journal} {Phys. Rev. B}\ }\textbf {\bibinfo {volume} {89}},\ \bibinfo
  {pages} {235127} (\bibinfo {year} {2014})}\BibitemShut {NoStop}%
\bibitem [{\citenamefont {Zhao}\ and\ \citenamefont {Lu}(2017)}]{ZhaoYX2}%
  \BibitemOpen
  \bibfield  {author} {\bibinfo {author} {\bibfnamefont {Y.~X.}\ \bibnamefont
  {Zhao}}\ and\ \bibinfo {author} {\bibfnamefont {Y.}~\bibnamefont {Lu}},\
  }\bibfield  {title} {\bibinfo {title} {{$PT$-Symmetric Real Dirac Fermions
  and Semimetals}},\ }\href {https://doi.org/10.1103/PhysRevLett.118.056401}
  {\bibfield  {journal} {\bibinfo  {journal} {Phys. Rev. Lett.}\ }\textbf
  {\bibinfo {volume} {118}},\ \bibinfo {pages} {056401} (\bibinfo {year}
  {2017})}\BibitemShut {NoStop}%
\bibitem [{\citenamefont {Chen}\ \emph {et~al.}(2022)\citenamefont {Chen},
  \citenamefont {Zeng}, \citenamefont {Chen}, \citenamefont {Zhao},
  \citenamefont {Sheng},\ and\ \citenamefont {Yang}}]{ChenCong}%
  \BibitemOpen
  \bibfield  {author} {\bibinfo {author} {\bibfnamefont {C.}~\bibnamefont
  {Chen}}, \bibinfo {author} {\bibfnamefont {X.-T.}\ \bibnamefont {Zeng}},
  \bibinfo {author} {\bibfnamefont {Z.}~\bibnamefont {Chen}}, \bibinfo {author}
  {\bibfnamefont {Y.~X.}\ \bibnamefont {Zhao}}, \bibinfo {author}
  {\bibfnamefont {X.-L.}\ \bibnamefont {Sheng}},\ and\ \bibinfo {author}
  {\bibfnamefont {S.~A.}\ \bibnamefont {Yang}},\ }\bibfield  {title} {\bibinfo
  {title} {{Second-Order Real Nodal-Line Semimetal in Three-Dimensional
  Graphdiyne}},\ }\href {https://doi.org/10.1103/PhysRevLett.128.026405}
  {\bibfield  {journal} {\bibinfo  {journal} {Phys. Rev. Lett.}\ }\textbf
  {\bibinfo {volume} {128}},\ \bibinfo {pages} {026405} (\bibinfo {year}
  {2022})}\BibitemShut {NoStop}%
\bibitem [{\citenamefont {Gao}\ \emph {et~al.}(2018)\citenamefont {Gao},
  \citenamefont {Zhu}, \citenamefont {Yi}, \citenamefont {Zhou}, \citenamefont
  {Zhang}, \citenamefont {Yin}, \citenamefont {Ding}, \citenamefont {Zhang},
  \citenamefont {Yi}, \citenamefont {Wang}, \citenamefont {Tong}, \citenamefont
  {Han}, \citenamefont {Liu},\ and\ \citenamefont {Zhang}}]{XGAO}%
  \BibitemOpen
  \bibfield  {author} {\bibinfo {author} {\bibfnamefont {X.}~\bibnamefont
  {Gao}}, \bibinfo {author} {\bibfnamefont {Y.}~\bibnamefont {Zhu}}, \bibinfo
  {author} {\bibfnamefont {D.}~\bibnamefont {Yi}}, \bibinfo {author}
  {\bibfnamefont {J.}~\bibnamefont {Zhou}}, \bibinfo {author} {\bibfnamefont
  {S.}~\bibnamefont {Zhang}}, \bibinfo {author} {\bibfnamefont
  {C.}~\bibnamefont {Yin}}, \bibinfo {author} {\bibfnamefont {F.}~\bibnamefont
  {Ding}}, \bibinfo {author} {\bibfnamefont {S.}~\bibnamefont {Zhang}},
  \bibinfo {author} {\bibfnamefont {X.}~\bibnamefont {Yi}}, \bibinfo {author}
  {\bibfnamefont {J.}~\bibnamefont {Wang}}, \bibinfo {author} {\bibfnamefont
  {L.}~\bibnamefont {Tong}}, \bibinfo {author} {\bibfnamefont {Y.}~\bibnamefont
  {Han}}, \bibinfo {author} {\bibfnamefont {Z.}~\bibnamefont {Liu}},\ and\
  \bibinfo {author} {\bibfnamefont {J.}~\bibnamefont {Zhang}},\ }\bibfield
  {title} {\bibinfo {title} {{Ultrathin graphdiyne film on graphene through
  solution-phase van der Waals epitaxy}},\ }\href
  {https://doi.org/10.1126/sciadv.aat6378} {\bibfield  {journal} {\bibinfo
  {journal} {Sci. Adv.}\ }\textbf {\bibinfo {volume} {4}},\ \bibinfo {pages}
  {eaat6378} (\bibinfo {year} {2018})}\BibitemShut {NoStop}%
\bibitem [{\citenamefont {Nomura}\ \emph {et~al.}(2018)\citenamefont {Nomura},
  \citenamefont {Habe}, \citenamefont {Sakamoto},\ and\ \citenamefont
  {Koshino}}]{Nomura}%
  \BibitemOpen
  \bibfield  {author} {\bibinfo {author} {\bibfnamefont {T.}~\bibnamefont
  {Nomura}}, \bibinfo {author} {\bibfnamefont {T.}~\bibnamefont {Habe}},
  \bibinfo {author} {\bibfnamefont {R.}~\bibnamefont {Sakamoto}},\ and\
  \bibinfo {author} {\bibfnamefont {M.}~\bibnamefont {Koshino}},\ }\bibfield
  {title} {\bibinfo {title} {{Three-dimensional graphdiyne as a topological
  nodal-line semimetal}},\ }\href
  {https://doi.org/10.1103/PhysRevMaterials.2.054204} {\bibfield  {journal}
  {\bibinfo  {journal} {Phys. Rev. Mater.}\ }\textbf {\bibinfo {volume} {2}},\
  \bibinfo {pages} {054204} (\bibinfo {year} {2018})}\BibitemShut {NoStop}%
\bibitem [{\citenamefont {Han}\ \emph {et~al.}(2024)\citenamefont {Han},
  \citenamefont {Liu}, \citenamefont {Cui}, \citenamefont {Liu},\ and\
  \citenamefont {Yu}}]{Yilin}%
  \BibitemOpen
  \bibfield  {author} {\bibinfo {author} {\bibfnamefont {Y.}~\bibnamefont
  {Han}}, \bibinfo {author} {\bibfnamefont {Y.}~\bibnamefont {Liu}}, \bibinfo
  {author} {\bibfnamefont {C.}~\bibnamefont {Cui}}, \bibinfo {author}
  {\bibfnamefont {C.-C.}\ \bibnamefont {Liu}},\ and\ \bibinfo {author}
  {\bibfnamefont {Z.-M.}\ \bibnamefont {Yu}},\ }\bibfield  {title} {\bibinfo
  {title} {{Crossed real nodal-line phonons in gold monobromide}},\ }\href
  {https://doi.org/10.1103/PhysRevB.110.184303} {\bibfield  {journal} {\bibinfo
   {journal} {Phys. Rev. B}\ }\textbf {\bibinfo {volume} {110}},\ \bibinfo
  {pages} {184303} (\bibinfo {year} {2024})}\BibitemShut {NoStop}%
\bibitem [{\citenamefont {Wang}\ \emph {et~al.}(2024)\citenamefont {Wang},
  \citenamefont {Bai}, \citenamefont {Wang}, \citenamefont {Cheng},
  \citenamefont {Qian}, \citenamefont {Wang}, \citenamefont {Zhang},
  \citenamefont {Yu},\ and\ \citenamefont {Yao}}]{XTwang}%
  \BibitemOpen
  \bibfield  {author} {\bibinfo {author} {\bibfnamefont {X.}~\bibnamefont
  {Wang}}, \bibinfo {author} {\bibfnamefont {J.}~\bibnamefont {Bai}}, \bibinfo
  {author} {\bibfnamefont {J.}~\bibnamefont {Wang}}, \bibinfo {author}
  {\bibfnamefont {Z.}~\bibnamefont {Cheng}}, \bibinfo {author} {\bibfnamefont
  {S.}~\bibnamefont {Qian}}, \bibinfo {author} {\bibfnamefont {W.}~\bibnamefont
  {Wang}}, \bibinfo {author} {\bibfnamefont {G.}~\bibnamefont {Zhang}},
  \bibinfo {author} {\bibfnamefont {Z.-M.}\ \bibnamefont {Yu}},\ and\ \bibinfo
  {author} {\bibfnamefont {Y.}~\bibnamefont {Yao}},\ }\bibfield  {title}
  {\bibinfo {title} {{Real Topological Phonons in 3D Carbon Allotropes}},\
  }\href {https://doi.org/10.1002/adma.202407437} {\bibfield  {journal}
  {\bibinfo  {journal} {Adv. Mater.}\ }\textbf {\bibinfo {volume} {36}},\
  \bibinfo {pages} {2407437} (\bibinfo {year} {2024})}\BibitemShut {NoStop}%
\bibitem [{\citenamefont {Li}\ \emph {et~al.}(2025{\natexlab{a}})\citenamefont
  {Li}, \citenamefont {Qian},\ and\ \citenamefont {Liu}}]{YongpanLi}%
  \BibitemOpen
  \bibfield  {author} {\bibinfo {author} {\bibfnamefont {Y.}~\bibnamefont
  {Li}}, \bibinfo {author} {\bibfnamefont {S.}~\bibnamefont {Qian}},\ and\
  \bibinfo {author} {\bibfnamefont {C.-C.}\ \bibnamefont {Liu}},\ }\bibfield
  {title} {\bibinfo {title} {{General construction of three-dimensional
  ${\mathbb{Z}}_{2}$ monopole charge nodal line semimetals and prediction of
  abundant candidate materials}},\ }\href
  {https://doi.org/10.1103/PhysRevB.111.125101} {\bibfield  {journal} {\bibinfo
   {journal} {Phys. Rev. B}\ }\textbf {\bibinfo {volume} {111}},\ \bibinfo
  {pages} {125101} (\bibinfo {year} {2025}{\natexlab{a}})}\BibitemShut
  {NoStop}%
\bibitem [{\citenamefont {Xue}\ \emph {et~al.}(2023)\citenamefont {Xue},
  \citenamefont {Chen}, \citenamefont {Cheng}, \citenamefont {Dai},
  \citenamefont {Long}, \citenamefont {Zhao},\ and\ \citenamefont
  {Zhang}}]{Haoran}%
  \BibitemOpen
  \bibfield  {author} {\bibinfo {author} {\bibfnamefont {H.}~\bibnamefont
  {Xue}}, \bibinfo {author} {\bibfnamefont {Z.~Y.}\ \bibnamefont {Chen}},
  \bibinfo {author} {\bibfnamefont {Z.}~\bibnamefont {Cheng}}, \bibinfo
  {author} {\bibfnamefont {J.~X.}\ \bibnamefont {Dai}}, \bibinfo {author}
  {\bibfnamefont {Y.}~\bibnamefont {Long}}, \bibinfo {author} {\bibfnamefont
  {Y.~X.}\ \bibnamefont {Zhao}},\ and\ \bibinfo {author} {\bibfnamefont
  {B.}~\bibnamefont {Zhang}},\ }\bibfield  {title} {\bibinfo {title}
  {{Stiefel-Whitney topological charges in a three-dimensional acoustic
  nodal-line crystal}},\ }\href {https://doi.org/10.1038/s41467-023-40252-7}
  {\bibfield  {journal} {\bibinfo  {journal} {Nat. Commun.}\ }\textbf {\bibinfo
  {volume} {14}},\ \bibinfo {pages} {4563} (\bibinfo {year}
  {2023})}\BibitemShut {NoStop}%
\bibitem [{\citenamefont {Xiang}\ \emph {et~al.}(2024)\citenamefont {Xiang},
  \citenamefont {Peng}, \citenamefont {Gao}, \citenamefont {Wu},\ and\
  \citenamefont {Wu}}]{PWu}%
  \BibitemOpen
  \bibfield  {author} {\bibinfo {author} {\bibfnamefont {X.}~\bibnamefont
  {Xiang}}, \bibinfo {author} {\bibfnamefont {Y.-G.}\ \bibnamefont {Peng}},
  \bibinfo {author} {\bibfnamefont {F.}~\bibnamefont {Gao}}, \bibinfo {author}
  {\bibfnamefont {X.}~\bibnamefont {Wu}},\ and\ \bibinfo {author}
  {\bibfnamefont {P.}~\bibnamefont {Wu}},\ }\bibfield  {title} {\bibinfo
  {title} {{Demonstration of Acoustic Higher-Order Topological Stiefel-Whitney
  Semimetal}},\ }\href {https://doi.org/10.1103/PhysRevLett.132.197202}
  {\bibfield  {journal} {\bibinfo  {journal} {Phys. Rev. Lett.}\ }\textbf
  {\bibinfo {volume} {132}},\ \bibinfo {pages} {197202} (\bibinfo {year}
  {2024})}\BibitemShut {NoStop}%
\bibitem [{\citenamefont {Salerno}\ \emph {et~al.}(2020)\citenamefont
  {Salerno}, \citenamefont {Goldman},\ and\ \citenamefont {Palumbo}}]{Salerno}%
  \BibitemOpen
  \bibfield  {author} {\bibinfo {author} {\bibfnamefont {G.}~\bibnamefont
  {Salerno}}, \bibinfo {author} {\bibfnamefont {N.}~\bibnamefont {Goldman}},\
  and\ \bibinfo {author} {\bibfnamefont {G.}~\bibnamefont {Palumbo}},\
  }\bibfield  {title} {\bibinfo {title} {{Floquet-engineering of nodal rings
  and nodal spheres and their characterization using the quantum metric}},\
  }\href {https://doi.org/10.1103/PhysRevResearch.2.013224} {\bibfield
  {journal} {\bibinfo  {journal} {Phys. Rev. Res.}\ }\textbf {\bibinfo {volume}
  {2}},\ \bibinfo {pages} {013224} (\bibinfo {year} {2020})}\BibitemShut
  {NoStop}%
\bibitem [{\citenamefont {Sun}\ \emph {et~al.}(2018)\citenamefont {Sun},
  \citenamefont {Zhang},\ and\ \citenamefont {Bzdu\ifmmode~\check{s}\else
  \v{s}\fi{}ek}}]{Tomas}%
  \BibitemOpen
  \bibfield  {author} {\bibinfo {author} {\bibfnamefont {X.-Q.}\ \bibnamefont
  {Sun}}, \bibinfo {author} {\bibfnamefont {S.-C.}\ \bibnamefont {Zhang}},\
  and\ \bibinfo {author} {\bibfnamefont {T.~c.~v.}\ \bibnamefont
  {Bzdu\ifmmode~\check{s}\else \v{s}\fi{}ek}},\ }\bibfield  {title} {\bibinfo
  {title} {Conversion rules for weyl points and nodal lines in topological
  media},\ }\href {https://doi.org/10.1103/PhysRevLett.121.106402} {\bibfield
  {journal} {\bibinfo  {journal} {Phys. Rev. Lett.}\ }\textbf {\bibinfo
  {volume} {121}},\ \bibinfo {pages} {106402} (\bibinfo {year}
  {2018})}\BibitemShut {NoStop}%
\bibitem [{\citenamefont {Lenggenhager}\ \emph {et~al.}(2021)\citenamefont
  {Lenggenhager}, \citenamefont {Liu}, \citenamefont {Tsirkin}, \citenamefont
  {Neupert},\ and\ \citenamefont {Bzdu\ifmmode~\check{s}\else
  \v{s}\fi{}ek}}]{Lenggenhager}%
  \BibitemOpen
  \bibfield  {author} {\bibinfo {author} {\bibfnamefont {P.~M.}\ \bibnamefont
  {Lenggenhager}}, \bibinfo {author} {\bibfnamefont {X.}~\bibnamefont {Liu}},
  \bibinfo {author} {\bibfnamefont {S.~S.}\ \bibnamefont {Tsirkin}}, \bibinfo
  {author} {\bibfnamefont {T.}~\bibnamefont {Neupert}},\ and\ \bibinfo {author}
  {\bibfnamefont {T.~c.~v.}\ \bibnamefont {Bzdu\ifmmode~\check{s}\else
  \v{s}\fi{}ek}},\ }\bibfield  {title} {\bibinfo {title} {From triple-point
  materials to multiband nodal links},\ }\href
  {https://doi.org/10.1103/PhysRevB.103.L121101} {\bibfield  {journal}
  {\bibinfo  {journal} {Phys. Rev. B}\ }\textbf {\bibinfo {volume} {103}},\
  \bibinfo {pages} {L121101} (\bibinfo {year} {2021})}\BibitemShut {NoStop}%
\bibitem [{\citenamefont {Wu}\ \emph {et~al.}(2024)\citenamefont {Wu},
  \citenamefont {Weng}, \citenamefont {Chi}, \citenamefont {Qi}, \citenamefont
  {Li}, \citenamefont {Zhao}, \citenamefont {Meng},\ and\ \citenamefont
  {Zhou}}]{WuMaopeng}%
  \BibitemOpen
  \bibfield  {author} {\bibinfo {author} {\bibfnamefont {M.}~\bibnamefont
  {Wu}}, \bibinfo {author} {\bibfnamefont {M.}~\bibnamefont {Weng}}, \bibinfo
  {author} {\bibfnamefont {Z.}~\bibnamefont {Chi}}, \bibinfo {author}
  {\bibfnamefont {Y.}~\bibnamefont {Qi}}, \bibinfo {author} {\bibfnamefont
  {H.}~\bibnamefont {Li}}, \bibinfo {author} {\bibfnamefont {Q.}~\bibnamefont
  {Zhao}}, \bibinfo {author} {\bibfnamefont {Y.}~\bibnamefont {Meng}},\ and\
  \bibinfo {author} {\bibfnamefont {J.}~\bibnamefont {Zhou}},\ }\bibfield
  {title} {\bibinfo {title} {Observing relative homotopic degeneracy
  conversions with circuit metamaterials},\ }\href
  {https://doi.org/10.1103/PhysRevLett.132.016605} {\bibfield  {journal}
  {\bibinfo  {journal} {Phys. Rev. Lett.}\ }\textbf {\bibinfo {volume} {132}},\
  \bibinfo {pages} {016605} (\bibinfo {year} {2024})}\BibitemShut {NoStop}%
\bibitem [{\citenamefont {Bouhon}\ \emph {et~al.}(2020)\citenamefont {Bouhon},
  \citenamefont {Wu}, \citenamefont {Slager}, \citenamefont {Weng},
  \citenamefont {Yazyev},\ and\ \citenamefont {Bzdu\ifmmode~\check{s}\else
  \v{s}\fi{}ek}}]{Tomas2}%
  \BibitemOpen
  \bibfield  {author} {\bibinfo {author} {\bibfnamefont {A.}~\bibnamefont
  {Bouhon}}, \bibinfo {author} {\bibfnamefont {Q.}~\bibnamefont {Wu}}, \bibinfo
  {author} {\bibfnamefont {R.-J.}\ \bibnamefont {Slager}}, \bibinfo {author}
  {\bibfnamefont {H.}~\bibnamefont {Weng}}, \bibinfo {author} {\bibfnamefont
  {O.~V.}\ \bibnamefont {Yazyev}},\ and\ \bibinfo {author} {\bibfnamefont
  {T.~c.~v.}\ \bibnamefont {Bzdu\ifmmode~\check{s}\else \v{s}\fi{}ek}},\
  }\bibfield  {title} {\bibinfo {title} {{Non-Abelian reciprocal braiding of
  Weyl points and its manifestation in ZrTe}},\ }\href
  {https://doi.org/10.1038/s41567-020-0967-9} {\bibfield  {journal} {\bibinfo
  {journal} {Nature Physics}\ }\textbf {\bibinfo {volume} {16}},\ \bibinfo
  {pages} {1137} (\bibinfo {year} {2020})}\BibitemShut {NoStop}%
\bibitem [{\citenamefont {Rudner}\ and\ \citenamefont {Lindner}(2020)}]{MarkS}%
  \BibitemOpen
  \bibfield  {author} {\bibinfo {author} {\bibfnamefont {M.~S.}\ \bibnamefont
  {Rudner}}\ and\ \bibinfo {author} {\bibfnamefont {N.~H.}\ \bibnamefont
  {Lindner}},\ }\bibfield  {title} {\bibinfo {title} {{Band structure
  engineering and non-equilibrium dynamics in Floquet topological
  insulators}},\ }\href {https://doi.org/10.1038/s42254-020-0170-z} {\bibfield
  {journal} {\bibinfo  {journal} {Nat. Rev. Phys.}\ }\textbf {\bibinfo {volume}
  {2}},\ \bibinfo {pages} {229} (\bibinfo {year} {2020})}\BibitemShut {NoStop}%
\bibitem [{\citenamefont {Oka}\ and\ \citenamefont
  {Kitamura}(2019)}]{Takashi2}%
  \BibitemOpen
  \bibfield  {author} {\bibinfo {author} {\bibfnamefont {T.}~\bibnamefont
  {Oka}}\ and\ \bibinfo {author} {\bibfnamefont {S.}~\bibnamefont {Kitamura}},\
  }\bibfield  {title} {\bibinfo {title} {{Floquet Engineering of Quantum
  Materials}},\ }\href
  {https://doi.org/10.1146/annurev-conmatphys-031218-013423} {\bibfield
  {journal} {\bibinfo  {journal} {Annu. Rev. Condens. Matter Phys.}\ }\textbf
  {\bibinfo {volume} {10}},\ \bibinfo {pages} {387} (\bibinfo {year}
  {2019})}\BibitemShut {NoStop}%
\bibitem [{\citenamefont {Bao}\ \emph {et~al.}(2022)\citenamefont {Bao},
  \citenamefont {Tang}, \citenamefont {Sun},\ and\ \citenamefont
  {Zhou}}]{ChanghuaBao}%
  \BibitemOpen
  \bibfield  {author} {\bibinfo {author} {\bibfnamefont {C.}~\bibnamefont
  {Bao}}, \bibinfo {author} {\bibfnamefont {P.}~\bibnamefont {Tang}}, \bibinfo
  {author} {\bibfnamefont {D.}~\bibnamefont {Sun}},\ and\ \bibinfo {author}
  {\bibfnamefont {S.}~\bibnamefont {Zhou}},\ }\bibfield  {title} {\bibinfo
  {title} {{Light-induced emergent phenomena in 2D materials and topological
  materials}},\ }\href {https://doi.org/10.1038/s42254-021-00388-1} {\bibfield
  {journal} {\bibinfo  {journal} {Nat. Rev. Phys.}\ }\textbf {\bibinfo {volume}
  {4}},\ \bibinfo {pages} {33} (\bibinfo {year} {2022})}\BibitemShut {NoStop}%
\bibitem [{\citenamefont {Wang}\ \emph
  {et~al.}(2013{\natexlab{b}})\citenamefont {Wang}, \citenamefont {Steinberg},
  \citenamefont {Jarillo-Herrero},\ and\ \citenamefont {Gedik}}]{YHWANG}%
  \BibitemOpen
  \bibfield  {author} {\bibinfo {author} {\bibfnamefont {Y.~H.}\ \bibnamefont
  {Wang}}, \bibinfo {author} {\bibfnamefont {H.}~\bibnamefont {Steinberg}},
  \bibinfo {author} {\bibfnamefont {P.}~\bibnamefont {Jarillo-Herrero}},\ and\
  \bibinfo {author} {\bibfnamefont {N.}~\bibnamefont {Gedik}},\ }\bibfield
  {title} {\bibinfo {title} {{Observation of Floquet-Bloch States on the
  Surface of a Topological Insulator}},\ }\href
  {https://doi.org/10.1126/science.1239834} {\bibfield  {journal} {\bibinfo
  {journal} {Science}\ }\textbf {\bibinfo {volume} {342}},\ \bibinfo {pages}
  {453} (\bibinfo {year} {2013}{\natexlab{b}})}\BibitemShut {NoStop}%
\bibitem [{\citenamefont {Mahmood}\ \emph {et~al.}(2016)\citenamefont
  {Mahmood}, \citenamefont {Chan}, \citenamefont {Alpichshev}, \citenamefont
  {Gardner}, \citenamefont {Lee}, \citenamefont {Lee},\ and\ \citenamefont
  {Gedik}}]{Fahad}%
  \BibitemOpen
  \bibfield  {author} {\bibinfo {author} {\bibfnamefont {F.}~\bibnamefont
  {Mahmood}}, \bibinfo {author} {\bibfnamefont {C.-K.}\ \bibnamefont {Chan}},
  \bibinfo {author} {\bibfnamefont {Z.}~\bibnamefont {Alpichshev}}, \bibinfo
  {author} {\bibfnamefont {D.}~\bibnamefont {Gardner}}, \bibinfo {author}
  {\bibfnamefont {Y.}~\bibnamefont {Lee}}, \bibinfo {author} {\bibfnamefont
  {P.~A.}\ \bibnamefont {Lee}},\ and\ \bibinfo {author} {\bibfnamefont
  {N.}~\bibnamefont {Gedik}},\ }\bibfield  {title} {\bibinfo {title}
  {{Selective scattering between Floquet–Bloch and Volkov states in a
  topological insulator}},\ }\href {https://doi.org/10.1038/nphys3609}
  {\bibfield  {journal} {\bibinfo  {journal} {Nat. Phys.}\ }\textbf {\bibinfo
  {volume} {12}},\ \bibinfo {pages} {306} (\bibinfo {year} {2016})}\BibitemShut
  {NoStop}%
\bibitem [{\citenamefont {Sie}\ \emph {et~al.}(2015)\citenamefont {Sie},
  \citenamefont {McIver}, \citenamefont {Lee}, \citenamefont {Fu},
  \citenamefont {Kong},\ and\ \citenamefont {Gedik}}]{Edbert}%
  \BibitemOpen
  \bibfield  {author} {\bibinfo {author} {\bibfnamefont {E.~J.}\ \bibnamefont
  {Sie}}, \bibinfo {author} {\bibfnamefont {J.~W.}\ \bibnamefont {McIver}},
  \bibinfo {author} {\bibfnamefont {Y.-H.}\ \bibnamefont {Lee}}, \bibinfo
  {author} {\bibfnamefont {L.}~\bibnamefont {Fu}}, \bibinfo {author}
  {\bibfnamefont {J.}~\bibnamefont {Kong}},\ and\ \bibinfo {author}
  {\bibfnamefont {N.}~\bibnamefont {Gedik}},\ }\bibfield  {title} {\bibinfo
  {title} {{Valley-selective optical Stark effect in monolayer WS$_2$}},\
  }\href {https://doi.org/10.1038/nmat4156} {\bibfield  {journal} {\bibinfo
  {journal} {Nat. Mater.}\ }\textbf {\bibinfo {volume} {14}},\ \bibinfo {pages}
  {290} (\bibinfo {year} {2015})}\BibitemShut {NoStop}%
\bibitem [{\citenamefont {Wang}\ \emph {et~al.}(2014)\citenamefont {Wang},
  \citenamefont {Steinberg}, \citenamefont {Jarillo-Herrero},\ and\
  \citenamefont {Gedik}}]{YHWANG2}%
  \BibitemOpen
  \bibfield  {author} {\bibinfo {author} {\bibfnamefont {Y.~H.}\ \bibnamefont
  {Wang}}, \bibinfo {author} {\bibfnamefont {H.}~\bibnamefont {Steinberg}},
  \bibinfo {author} {\bibfnamefont {P.}~\bibnamefont {Jarillo-Herrero}},\ and\
  \bibinfo {author} {\bibfnamefont {N.}~\bibnamefont {Gedik}},\ }\bibfield
  {title} {\bibinfo {title} {{Ultrafast generation of pseudo-magnetic field for
  valley excitons in WSe2 monolayers}},\ }\href
  {https://doi.org/10.1126/science.1258122} {\bibfield  {journal} {\bibinfo
  {journal} {Science}\ }\textbf {\bibinfo {volume} {346}},\ \bibinfo {pages}
  {1205} (\bibinfo {year} {2014})}\BibitemShut {NoStop}%
\bibitem [{\citenamefont {Lindner}\ \emph {et~al.}(2011)\citenamefont
  {Lindner}, \citenamefont {Refael},\ and\ \citenamefont {Galitski}}]{Netanel}%
  \BibitemOpen
  \bibfield  {author} {\bibinfo {author} {\bibfnamefont {N.~H.}\ \bibnamefont
  {Lindner}}, \bibinfo {author} {\bibfnamefont {G.}~\bibnamefont {Refael}},\
  and\ \bibinfo {author} {\bibfnamefont {V.}~\bibnamefont {Galitski}},\
  }\bibfield  {title} {\bibinfo {title} {{Floquet topological insulator in
  semiconductor quantum wells}},\ }\href {https://doi.org/10.1038/nphys1926}
  {\bibfield  {journal} {\bibinfo  {journal} {Nat. Phys.}\ }\textbf {\bibinfo
  {volume} {7}},\ \bibinfo {pages} {490} (\bibinfo {year} {2011})}\BibitemShut
  {NoStop}%
\bibitem [{\citenamefont {Titum}\ \emph {et~al.}(2017)\citenamefont {Titum},
  \citenamefont {Lindner},\ and\ \citenamefont {Refael}}]{Titum}%
  \BibitemOpen
  \bibfield  {author} {\bibinfo {author} {\bibfnamefont {P.}~\bibnamefont
  {Titum}}, \bibinfo {author} {\bibfnamefont {N.~H.}\ \bibnamefont {Lindner}},\
  and\ \bibinfo {author} {\bibfnamefont {G.}~\bibnamefont {Refael}},\
  }\bibfield  {title} {\bibinfo {title} {{Disorder-induced transitions in
  resonantly driven Floquet topological insulators}},\ }\href
  {https://doi.org/10.1103/PhysRevB.96.054207} {\bibfield  {journal} {\bibinfo
  {journal} {Phys. Rev. B}\ }\textbf {\bibinfo {volume} {96}},\ \bibinfo
  {pages} {054207} (\bibinfo {year} {2017})}\BibitemShut {NoStop}%
\bibitem [{\citenamefont {Titum}\ \emph {et~al.}(2015)\citenamefont {Titum},
  \citenamefont {Lindner}, \citenamefont {Rechtsman},\ and\ \citenamefont
  {Refael}}]{Titum2}%
  \BibitemOpen
  \bibfield  {author} {\bibinfo {author} {\bibfnamefont {P.}~\bibnamefont
  {Titum}}, \bibinfo {author} {\bibfnamefont {N.~H.}\ \bibnamefont {Lindner}},
  \bibinfo {author} {\bibfnamefont {M.~C.}\ \bibnamefont {Rechtsman}},\ and\
  \bibinfo {author} {\bibfnamefont {G.}~\bibnamefont {Refael}},\ }\bibfield
  {title} {\bibinfo {title} {{Disorder-Induced Floquet Topological
  Insulators}},\ }\href {https://doi.org/10.1103/PhysRevLett.114.056801}
  {\bibfield  {journal} {\bibinfo  {journal} {Phys. Rev. Lett.}\ }\textbf
  {\bibinfo {volume} {114}},\ \bibinfo {pages} {056801} (\bibinfo {year}
  {2015})}\BibitemShut {NoStop}%
\bibitem [{\citenamefont {Oka}\ and\ \citenamefont {Aoki}(2009)}]{Oka}%
  \BibitemOpen
  \bibfield  {author} {\bibinfo {author} {\bibfnamefont {T.}~\bibnamefont
  {Oka}}\ and\ \bibinfo {author} {\bibfnamefont {H.}~\bibnamefont {Aoki}},\
  }\bibfield  {title} {\bibinfo {title} {{Photovoltaic Hall effect in
  graphene}},\ }\href {https://doi.org/10.1103/PhysRevB.79.081406} {\bibfield
  {journal} {\bibinfo  {journal} {Phys. Rev. B}\ }\textbf {\bibinfo {volume}
  {79}},\ \bibinfo {pages} {081406} (\bibinfo {year} {2009})}\BibitemShut
  {NoStop}%
\bibitem [{\citenamefont {Kitagawa}\ \emph {et~al.}(2011)\citenamefont
  {Kitagawa}, \citenamefont {Oka}, \citenamefont {Brataas}, \citenamefont
  {Fu},\ and\ \citenamefont {Demler}}]{FuLiang}%
  \BibitemOpen
  \bibfield  {author} {\bibinfo {author} {\bibfnamefont {T.}~\bibnamefont
  {Kitagawa}}, \bibinfo {author} {\bibfnamefont {T.}~\bibnamefont {Oka}},
  \bibinfo {author} {\bibfnamefont {A.}~\bibnamefont {Brataas}}, \bibinfo
  {author} {\bibfnamefont {L.}~\bibnamefont {Fu}},\ and\ \bibinfo {author}
  {\bibfnamefont {E.}~\bibnamefont {Demler}},\ }\bibfield  {title} {\bibinfo
  {title} {{Transport properties of nonequilibrium systems under the
  application of light: Photoinduced quantum Hall insulators without Landau
  levels}},\ }\href {https://doi.org/10.1103/PhysRevB.84.235108} {\bibfield
  {journal} {\bibinfo  {journal} {Phys. Rev. B}\ }\textbf {\bibinfo {volume}
  {84}},\ \bibinfo {pages} {235108} (\bibinfo {year} {2011})}\BibitemShut
  {NoStop}%
\bibitem [{\citenamefont {Usaj}\ \emph {et~al.}(2014)\citenamefont {Usaj},
  \citenamefont {Perez-Piskunow}, \citenamefont {Foa~Torres},\ and\
  \citenamefont {Balseiro}}]{Usaj}%
  \BibitemOpen
  \bibfield  {author} {\bibinfo {author} {\bibfnamefont {G.}~\bibnamefont
  {Usaj}}, \bibinfo {author} {\bibfnamefont {P.~M.}\ \bibnamefont
  {Perez-Piskunow}}, \bibinfo {author} {\bibfnamefont {L.~E.~F.}\ \bibnamefont
  {Foa~Torres}},\ and\ \bibinfo {author} {\bibfnamefont {C.~A.}\ \bibnamefont
  {Balseiro}},\ }\bibfield  {title} {\bibinfo {title} {{Irradiated graphene as
  a tunable Floquet topological insulator}},\ }\href
  {https://doi.org/10.1103/PhysRevB.90.115423} {\bibfield  {journal} {\bibinfo
  {journal} {Phys. Rev. B}\ }\textbf {\bibinfo {volume} {90}},\ \bibinfo
  {pages} {115423} (\bibinfo {year} {2014})}\BibitemShut {NoStop}%
\bibitem [{\citenamefont {Chan}\ \emph
  {et~al.}(2016{\natexlab{b}})\citenamefont {Chan}, \citenamefont {Oh},
  \citenamefont {Han},\ and\ \citenamefont {Lee}}]{Chan}%
  \BibitemOpen
  \bibfield  {author} {\bibinfo {author} {\bibfnamefont {C.-K.}\ \bibnamefont
  {Chan}}, \bibinfo {author} {\bibfnamefont {Y.-T.}\ \bibnamefont {Oh}},
  \bibinfo {author} {\bibfnamefont {J.~H.}\ \bibnamefont {Han}},\ and\ \bibinfo
  {author} {\bibfnamefont {P.~A.}\ \bibnamefont {Lee}},\ }\bibfield  {title}
  {\bibinfo {title} {{Type-II Weyl cone transitions in driven semimetals}},\
  }\href {https://doi.org/10.1103/PhysRevB.94.121106} {\bibfield  {journal}
  {\bibinfo  {journal} {Phys. Rev. B}\ }\textbf {\bibinfo {volume} {94}},\
  \bibinfo {pages} {121106} (\bibinfo {year} {2016}{\natexlab{b}})}\BibitemShut
  {NoStop}%
\bibitem [{\citenamefont {Chen}\ \emph {et~al.}(2018)\citenamefont {Chen},
  \citenamefont {Zhou},\ and\ \citenamefont {Xu}}]{ChenRui}%
  \BibitemOpen
  \bibfield  {author} {\bibinfo {author} {\bibfnamefont {R.}~\bibnamefont
  {Chen}}, \bibinfo {author} {\bibfnamefont {B.}~\bibnamefont {Zhou}},\ and\
  \bibinfo {author} {\bibfnamefont {D.-H.}\ \bibnamefont {Xu}},\ }\bibfield
  {title} {\bibinfo {title} {{Floquet Weyl semimetals in light-irradiated
  type-II and hybrid line-node semimetals}},\ }\href
  {https://doi.org/10.1103/PhysRevB.97.155152} {\bibfield  {journal} {\bibinfo
  {journal} {Phys. Rev. B}\ }\textbf {\bibinfo {volume} {97}},\ \bibinfo
  {pages} {155152} (\bibinfo {year} {2018})}\BibitemShut {NoStop}%
\bibitem [{\citenamefont {Yan}\ and\ \citenamefont {Wang}(2017)}]{Yan2}%
  \BibitemOpen
  \bibfield  {author} {\bibinfo {author} {\bibfnamefont {Z.}~\bibnamefont
  {Yan}}\ and\ \bibinfo {author} {\bibfnamefont {Z.}~\bibnamefont {Wang}},\
  }\bibfield  {title} {\bibinfo {title} {{Floquet multi-Weyl points in
  crossing-nodal-line semimetals}},\ }\href
  {https://doi.org/10.1103/PhysRevB.96.041206} {\bibfield  {journal} {\bibinfo
  {journal} {Phys. Rev. B}\ }\textbf {\bibinfo {volume} {96}},\ \bibinfo
  {pages} {041206} (\bibinfo {year} {2017})}\BibitemShut {NoStop}%
\bibitem [{\citenamefont {Deng}\ \emph {et~al.}(2020)\citenamefont {Deng},
  \citenamefont {Zheng}, \citenamefont {Zhan}, \citenamefont {Fan},
  \citenamefont {Wu},\ and\ \citenamefont {Wang}}]{DengTingwei}%
  \BibitemOpen
  \bibfield  {author} {\bibinfo {author} {\bibfnamefont {T.}~\bibnamefont
  {Deng}}, \bibinfo {author} {\bibfnamefont {B.}~\bibnamefont {Zheng}},
  \bibinfo {author} {\bibfnamefont {F.}~\bibnamefont {Zhan}}, \bibinfo {author}
  {\bibfnamefont {J.}~\bibnamefont {Fan}}, \bibinfo {author} {\bibfnamefont
  {X.}~\bibnamefont {Wu}},\ and\ \bibinfo {author} {\bibfnamefont
  {R.}~\bibnamefont {Wang}},\ }\bibfield  {title} {\bibinfo {title}
  {{Photoinduced Floquet mixed-Weyl semimetallic phase in a carbon
  allotrope}},\ }\href {https://doi.org/10.1103/PhysRevB.102.201105} {\bibfield
   {journal} {\bibinfo  {journal} {Phys. Rev. B}\ }\textbf {\bibinfo {volume}
  {102}},\ \bibinfo {pages} {201105} (\bibinfo {year} {2020})}\BibitemShut
  {NoStop}%
\bibitem [{\citenamefont {Hübener}\ \emph {et~al.}(2017)\citenamefont
  {Hübener}, \citenamefont {Sentef}, \citenamefont {Giovannini}, \citenamefont
  {Kemper},\ and\ \citenamefont {Rubio}}]{Hannes}%
  \BibitemOpen
  \bibfield  {author} {\bibinfo {author} {\bibfnamefont {H.}~\bibnamefont
  {Hübener}}, \bibinfo {author} {\bibfnamefont {M.~A.}\ \bibnamefont
  {Sentef}}, \bibinfo {author} {\bibfnamefont {U.~D.}\ \bibnamefont
  {Giovannini}}, \bibinfo {author} {\bibfnamefont {A.~F.}\ \bibnamefont
  {Kemper}},\ and\ \bibinfo {author} {\bibfnamefont {A.}~\bibnamefont
  {Rubio}},\ }\bibfield  {title} {\bibinfo {title} {{Creating stable
  Floquet–Weyl semimetals by laser-driving of 3D Dirac materials}},\ }\href
  {https://doi.org/10.1038/ncomms13940} {\bibfield  {journal} {\bibinfo
  {journal} {Nat. Commun.}\ }\textbf {\bibinfo {volume} {8}},\ \bibinfo {pages}
  {13940} (\bibinfo {year} {2017})}\BibitemShut {NoStop}%
\bibitem [{\citenamefont {Liu}\ \emph {et~al.}(2018)\citenamefont {Liu},
  \citenamefont {Sun}, \citenamefont {Cheng}, \citenamefont {Liu},\ and\
  \citenamefont {Meng}}]{LiuHang}%
  \BibitemOpen
  \bibfield  {author} {\bibinfo {author} {\bibfnamefont {H.}~\bibnamefont
  {Liu}}, \bibinfo {author} {\bibfnamefont {J.-T.}\ \bibnamefont {Sun}},
  \bibinfo {author} {\bibfnamefont {C.}~\bibnamefont {Cheng}}, \bibinfo
  {author} {\bibfnamefont {F.}~\bibnamefont {Liu}},\ and\ \bibinfo {author}
  {\bibfnamefont {S.}~\bibnamefont {Meng}},\ }\bibfield  {title} {\bibinfo
  {title} {{Photoinduced Nonequilibrium Topological States in Strained Black
  Phosphorus}},\ }\href {https://doi.org/10.1103/PhysRevLett.120.237403}
  {\bibfield  {journal} {\bibinfo  {journal} {Phys. Rev. Lett.}\ }\textbf
  {\bibinfo {volume} {120}},\ \bibinfo {pages} {237403} (\bibinfo {year}
  {2018})}\BibitemShut {NoStop}%
\bibitem [{\citenamefont {Yan}\ \emph {et~al.}(2017)\citenamefont {Yan},
  \citenamefont {Bi}, \citenamefont {Shen}, \citenamefont {Lu}, \citenamefont
  {Zhang},\ and\ \citenamefont {Wang}}]{YanZhongbo2}%
  \BibitemOpen
  \bibfield  {author} {\bibinfo {author} {\bibfnamefont {Z.}~\bibnamefont
  {Yan}}, \bibinfo {author} {\bibfnamefont {R.}~\bibnamefont {Bi}}, \bibinfo
  {author} {\bibfnamefont {H.}~\bibnamefont {Shen}}, \bibinfo {author}
  {\bibfnamefont {L.}~\bibnamefont {Lu}}, \bibinfo {author} {\bibfnamefont
  {S.-C.}\ \bibnamefont {Zhang}},\ and\ \bibinfo {author} {\bibfnamefont
  {Z.}~\bibnamefont {Wang}},\ }\bibfield  {title} {\bibinfo {title}
  {{Nodal-link semimetals}},\ }\href
  {https://doi.org/10.1103/PhysRevB.96.041103} {\bibfield  {journal} {\bibinfo
  {journal} {Phys. Rev. B}\ }\textbf {\bibinfo {volume} {96}},\ \bibinfo
  {pages} {041103} (\bibinfo {year} {2017})}\BibitemShut {NoStop}%
\bibitem [{\citenamefont {Narayan}(2016)}]{Awa}%
  \BibitemOpen
  \bibfield  {author} {\bibinfo {author} {\bibfnamefont {A.}~\bibnamefont
  {Narayan}},\ }\bibfield  {title} {\bibinfo {title} {{Tunable point nodes from
  line-node semimetals via application of light}},\ }\href
  {https://doi.org/10.1103/PhysRevB.94.041409} {\bibfield  {journal} {\bibinfo
  {journal} {Phys. Rev. B}\ }\textbf {\bibinfo {volume} {94}},\ \bibinfo
  {pages} {041409} (\bibinfo {year} {2016})}\BibitemShut {NoStop}%
\bibitem [{\citenamefont {Ghosh}\ \emph {et~al.}(2020)\citenamefont {Ghosh},
  \citenamefont {Paul},\ and\ \citenamefont {Saha}}]{Ghosh}%
  \BibitemOpen
  \bibfield  {author} {\bibinfo {author} {\bibfnamefont {A.~K.}\ \bibnamefont
  {Ghosh}}, \bibinfo {author} {\bibfnamefont {G.~C.}\ \bibnamefont {Paul}},\
  and\ \bibinfo {author} {\bibfnamefont {A.}~\bibnamefont {Saha}},\ }\bibfield
  {title} {\bibinfo {title} {{Higher order topological insulator via periodic
  driving}},\ }\href {https://doi.org/10.1103/PhysRevB.101.235403} {\bibfield
  {journal} {\bibinfo  {journal} {Phys. Rev. B}\ }\textbf {\bibinfo {volume}
  {101}},\ \bibinfo {pages} {235403} (\bibinfo {year} {2020})}\BibitemShut
  {NoStop}%
\bibitem [{\citenamefont {Rodriguez-Vega}\ \emph {et~al.}(2019)\citenamefont
  {Rodriguez-Vega}, \citenamefont {Kumar},\ and\ \citenamefont
  {Seradjeh}}]{Rodriguez}%
  \BibitemOpen
  \bibfield  {author} {\bibinfo {author} {\bibfnamefont {M.}~\bibnamefont
  {Rodriguez-Vega}}, \bibinfo {author} {\bibfnamefont {A.}~\bibnamefont
  {Kumar}},\ and\ \bibinfo {author} {\bibfnamefont {B.}~\bibnamefont
  {Seradjeh}},\ }\bibfield  {title} {\bibinfo {title} {{Higher-order Floquet
  topological phases with corner and bulk bound states}},\ }\href
  {https://doi.org/10.1103/PhysRevB.100.085138} {\bibfield  {journal} {\bibinfo
   {journal} {Phys. Rev. B}\ }\textbf {\bibinfo {volume} {100}},\ \bibinfo
  {pages} {085138} (\bibinfo {year} {2019})}\BibitemShut {NoStop}%
\bibitem [{\citenamefont {Huang}\ and\ \citenamefont {Liu}(2020)}]{Huang}%
  \BibitemOpen
  \bibfield  {author} {\bibinfo {author} {\bibfnamefont {B.}~\bibnamefont
  {Huang}}\ and\ \bibinfo {author} {\bibfnamefont {W.~V.}\ \bibnamefont
  {Liu}},\ }\bibfield  {title} {\bibinfo {title} {{Floquet Higher-Order
  Topological Insulators with Anomalous Dynamical Polarization}},\ }\href
  {https://doi.org/10.1103/PhysRevLett.124.216601} {\bibfield  {journal}
  {\bibinfo  {journal} {Phys. Rev. Lett.}\ }\textbf {\bibinfo {volume} {124}},\
  \bibinfo {pages} {216601} (\bibinfo {year} {2020})}\BibitemShut {NoStop}%
\bibitem [{\citenamefont {Zhu}\ \emph {et~al.}(2021)\citenamefont {Zhu},
  \citenamefont {Chong},\ and\ \citenamefont {Gong}}]{ZhuWeiwei}%
  \BibitemOpen
  \bibfield  {author} {\bibinfo {author} {\bibfnamefont {W.}~\bibnamefont
  {Zhu}}, \bibinfo {author} {\bibfnamefont {Y.~D.}\ \bibnamefont {Chong}},\
  and\ \bibinfo {author} {\bibfnamefont {J.}~\bibnamefont {Gong}},\ }\bibfield
  {title} {\bibinfo {title} {{Symmetry analysis of anomalous Floquet
  topological phases}},\ }\href {https://doi.org/10.1103/PhysRevB.104.L020302}
  {\bibfield  {journal} {\bibinfo  {journal} {Phys. Rev. B}\ }\textbf {\bibinfo
  {volume} {104}},\ \bibinfo {pages} {L020302} (\bibinfo {year}
  {2021})}\BibitemShut {NoStop}%
\bibitem [{\citenamefont {Du}\ \emph {et~al.}(2022)\citenamefont {Du},
  \citenamefont {Chen}, \citenamefont {Wang},\ and\ \citenamefont
  {Xu}}]{DuXiuLi}%
  \BibitemOpen
  \bibfield  {author} {\bibinfo {author} {\bibfnamefont {X.-L.}\ \bibnamefont
  {Du}}, \bibinfo {author} {\bibfnamefont {R.}~\bibnamefont {Chen}}, \bibinfo
  {author} {\bibfnamefont {R.}~\bibnamefont {Wang}},\ and\ \bibinfo {author}
  {\bibfnamefont {D.-H.}\ \bibnamefont {Xu}},\ }\bibfield  {title} {\bibinfo
  {title} {{Weyl nodes with higher-order topology in an optically driven
  nodal-line semimetal}},\ }\href
  {https://doi.org/10.1103/PhysRevB.105.L081102} {\bibfield  {journal}
  {\bibinfo  {journal} {Phys. Rev. B}\ }\textbf {\bibinfo {volume} {105}},\
  \bibinfo {pages} {L081102} (\bibinfo {year} {2022})}\BibitemShut {NoStop}%
\bibitem [{\citenamefont {Wang}\ \emph {et~al.}(2023)\citenamefont {Wang},
  \citenamefont {Wang}, \citenamefont {Sun}, \citenamefont {Chen},\ and\
  \citenamefont {Xu}}]{ZMWang}%
  \BibitemOpen
  \bibfield  {author} {\bibinfo {author} {\bibfnamefont {Z.-M.}\ \bibnamefont
  {Wang}}, \bibinfo {author} {\bibfnamefont {R.}~\bibnamefont {Wang}}, \bibinfo
  {author} {\bibfnamefont {J.-H.}\ \bibnamefont {Sun}}, \bibinfo {author}
  {\bibfnamefont {T.-Y.}\ \bibnamefont {Chen}},\ and\ \bibinfo {author}
  {\bibfnamefont {D.-H.}\ \bibnamefont {Xu}},\ }\bibfield  {title} {\bibinfo
  {title} {Floquet weyl semimetal phases in light-irradiated higher-order
  topological dirac semimetals},\ }\href
  {https://doi.org/10.1103/PhysRevB.107.L121407} {\bibfield  {journal}
  {\bibinfo  {journal} {Phys. Rev. B}\ }\textbf {\bibinfo {volume} {107}},\
  \bibinfo {pages} {L121407} (\bibinfo {year} {2023})}\BibitemShut {NoStop}%
\bibitem [{\citenamefont {Wang}\ \emph {et~al.}(2021)\citenamefont {Wang},
  \citenamefont {Wu},\ and\ \citenamefont {An}}]{BQWang}%
  \BibitemOpen
  \bibfield  {author} {\bibinfo {author} {\bibfnamefont {B.-Q.}\ \bibnamefont
  {Wang}}, \bibinfo {author} {\bibfnamefont {H.}~\bibnamefont {Wu}},\ and\
  \bibinfo {author} {\bibfnamefont {J.-H.}\ \bibnamefont {An}},\ }\bibfield
  {title} {\bibinfo {title} {Engineering exotic second-order topological
  semimetals by periodic driving},\ }\href
  {https://doi.org/10.1103/PhysRevB.104.205117} {\bibfield  {journal} {\bibinfo
   {journal} {Phys. Rev. B}\ }\textbf {\bibinfo {volume} {104}},\ \bibinfo
  {pages} {205117} (\bibinfo {year} {2021})}\BibitemShut {NoStop}%
\bibitem [{\citenamefont {Ghosh}\ \emph {et~al.}(2022)\citenamefont {Ghosh},
  \citenamefont {Saha},\ and\ \citenamefont {Sengupta}}]{GhoshWSM}%
  \BibitemOpen
  \bibfield  {author} {\bibinfo {author} {\bibfnamefont {S.}~\bibnamefont
  {Ghosh}}, \bibinfo {author} {\bibfnamefont {K.}~\bibnamefont {Saha}},\ and\
  \bibinfo {author} {\bibfnamefont {K.}~\bibnamefont {Sengupta}},\ }\bibfield
  {title} {\bibinfo {title} {Hinge-mode dynamics of periodically driven
  higher-order weyl semimetals},\ }\href
  {https://doi.org/10.1103/PhysRevB.105.224312} {\bibfield  {journal} {\bibinfo
   {journal} {Phys. Rev. B}\ }\textbf {\bibinfo {volume} {105}},\ \bibinfo
  {pages} {224312} (\bibinfo {year} {2022})}\BibitemShut {NoStop}%
\bibitem [{\citenamefont {Taguchi}\ \emph {et~al.}(2016)\citenamefont
  {Taguchi}, \citenamefont {Xu}, \citenamefont {Yamakage},\ and\ \citenamefont
  {Law}}]{DHXu}%
  \BibitemOpen
  \bibfield  {author} {\bibinfo {author} {\bibfnamefont {K.}~\bibnamefont
  {Taguchi}}, \bibinfo {author} {\bibfnamefont {D.-H.}\ \bibnamefont {Xu}},
  \bibinfo {author} {\bibfnamefont {A.}~\bibnamefont {Yamakage}},\ and\
  \bibinfo {author} {\bibfnamefont {K.~T.}\ \bibnamefont {Law}},\ }\bibfield
  {title} {\bibinfo {title} {{Photovoltaic anomalous Hall effect in line-node
  semimetals}},\ }\href {https://doi.org/10.1103/PhysRevB.94.155206} {\bibfield
   {journal} {\bibinfo  {journal} {Phys. Rev. B}\ }\textbf {\bibinfo {volume}
  {94}},\ \bibinfo {pages} {155206} (\bibinfo {year} {2016})}\BibitemShut
  {NoStop}%
\bibitem [{\citenamefont {Chan}\ \emph
  {et~al.}(2016{\natexlab{c}})\citenamefont {Chan}, \citenamefont {Lee},
  \citenamefont {Burch}, \citenamefont {Han},\ and\ \citenamefont
  {Ran}}]{Chan2}%
  \BibitemOpen
  \bibfield  {author} {\bibinfo {author} {\bibfnamefont {C.-K.}\ \bibnamefont
  {Chan}}, \bibinfo {author} {\bibfnamefont {P.~A.}\ \bibnamefont {Lee}},
  \bibinfo {author} {\bibfnamefont {K.~S.}\ \bibnamefont {Burch}}, \bibinfo
  {author} {\bibfnamefont {J.~H.}\ \bibnamefont {Han}},\ and\ \bibinfo {author}
  {\bibfnamefont {Y.}~\bibnamefont {Ran}},\ }\bibfield  {title} {\bibinfo
  {title} {{When Chiral Photons Meet Chiral Fermions: Photoinduced Anomalous
  Hall Effects in Weyl Semimetals}},\ }\href
  {https://doi.org/10.1103/PhysRevLett.116.026805} {\bibfield  {journal}
  {\bibinfo  {journal} {Phys. Rev. Lett.}\ }\textbf {\bibinfo {volume} {116}},\
  \bibinfo {pages} {026805} (\bibinfo {year} {2016}{\natexlab{c}})}\BibitemShut
  {NoStop}%
\bibitem [{\citenamefont {Gupta}()}]{Amit}%
  \BibitemOpen
  \bibfield  {author} {\bibinfo {author} {\bibfnamefont {A.}~\bibnamefont
  {Gupta}},\ }\bibfield  {title} {\bibinfo {title} {{Floquet dynamics in
  multi-Weyl semimetals}},\ }\href {https://doi.org/10.48550/arXiv.1703.07271}
  {\bibinfo  {journal} {arXiv:1703.07271}\ }\BibitemShut {NoStop}%
\bibitem [{\citenamefont {Xu}\ \emph {et~al.}(2021)\citenamefont {Xu},
  \citenamefont {Zhou},\ and\ \citenamefont {Li}}]{Haowei}%
  \BibitemOpen
\bibfield  {journal} {  }\bibfield  {author} {\bibinfo {author} {\bibfnamefont
  {H.}~\bibnamefont {Xu}}, \bibinfo {author} {\bibfnamefont {J.}~\bibnamefont
  {Zhou}},\ and\ \bibinfo {author} {\bibfnamefont {J.}~\bibnamefont {Li}},\
  }\bibfield  {title} {\bibinfo {title} {{Light-Induced Quantum Anomalous Hall
  Effect on the 2D Surfaces of 3D Topological Insulators}},\ }\href
  {https://doi.org/10.1002/advs.202101508} {\bibfield  {journal} {\bibinfo
  {journal} {Adv. Sci.}\ }\textbf {\bibinfo {volume} {8}},\ \bibinfo {pages}
  {2101508} (\bibinfo {year} {2021})}\BibitemShut {NoStop}%
\bibitem [{\citenamefont {Ning}\ \emph {et~al.}(2022)\citenamefont {Ning},
  \citenamefont {Zheng}, \citenamefont {Xu},\ and\ \citenamefont
  {Wang}}]{DHXu4}%
  \BibitemOpen
  \bibfield  {author} {\bibinfo {author} {\bibfnamefont {Z.}~\bibnamefont
  {Ning}}, \bibinfo {author} {\bibfnamefont {B.}~\bibnamefont {Zheng}},
  \bibinfo {author} {\bibfnamefont {D.-H.}\ \bibnamefont {Xu}},\ and\ \bibinfo
  {author} {\bibfnamefont {R.}~\bibnamefont {Wang}},\ }\bibfield  {title}
  {\bibinfo {title} {Photoinduced quantum anomalous hall states in the
  topological anderson insulator},\ }\href
  {https://doi.org/10.1103/PhysRevB.105.035103} {\bibfield  {journal} {\bibinfo
   {journal} {Phys. Rev. B}\ }\textbf {\bibinfo {volume} {105}},\ \bibinfo
  {pages} {035103} (\bibinfo {year} {2022})}\BibitemShut {NoStop}%
\bibitem [{\citenamefont {Yan}\ and\ \citenamefont {Wang}(2016)}]{YanZhongbo}%
  \BibitemOpen
  \bibfield  {author} {\bibinfo {author} {\bibfnamefont {Z.}~\bibnamefont
  {Yan}}\ and\ \bibinfo {author} {\bibfnamefont {Z.}~\bibnamefont {Wang}},\
  }\bibfield  {title} {\bibinfo {title} {{Tunable Weyl Points in Periodically
  Driven Nodal Line Semimetals}},\ }\href
  {https://doi.org/10.1103/PhysRevLett.117.087402} {\bibfield  {journal}
  {\bibinfo  {journal} {Phys. Rev. Lett.}\ }\textbf {\bibinfo {volume} {117}},\
  \bibinfo {pages} {087402} (\bibinfo {year} {2016})}\BibitemShut {NoStop}%
\bibitem [{\citenamefont {Menon}\ \emph {et~al.}(2018)\citenamefont {Menon},
  \citenamefont {Chowdhury},\ and\ \citenamefont {Basu}}]{Menon}%
  \BibitemOpen
  \bibfield  {author} {\bibinfo {author} {\bibfnamefont {A.}~\bibnamefont
  {Menon}}, \bibinfo {author} {\bibfnamefont {D.}~\bibnamefont {Chowdhury}},\
  and\ \bibinfo {author} {\bibfnamefont {B.}~\bibnamefont {Basu}},\ }\bibfield
  {title} {\bibinfo {title} {{Photoinduced tunable anomalous Hall and Nernst
  effects in tilted Weyl semimetals using Floquet theory}},\ }\href
  {https://doi.org/10.1103/PhysRevB.98.205109} {\bibfield  {journal} {\bibinfo
  {journal} {Phys. Rev. B}\ }\textbf {\bibinfo {volume} {98}},\ \bibinfo
  {pages} {205109} (\bibinfo {year} {2018})}\BibitemShut {NoStop}%
\bibitem [{\citenamefont {Nag}\ \emph {et~al.}(2020)\citenamefont {Nag},
  \citenamefont {Menon},\ and\ \citenamefont {Basu}}]{Nag}%
  \BibitemOpen
  \bibfield  {author} {\bibinfo {author} {\bibfnamefont {T.}~\bibnamefont
  {Nag}}, \bibinfo {author} {\bibfnamefont {A.}~\bibnamefont {Menon}},\ and\
  \bibinfo {author} {\bibfnamefont {B.}~\bibnamefont {Basu}},\ }\bibfield
  {title} {\bibinfo {title} {{Thermoelectric transport properties of Floquet
  multi-Weyl semimetals}},\ }\href
  {https://doi.org/10.1103/PhysRevB.102.014307} {\bibfield  {journal} {\bibinfo
   {journal} {Phys. Rev. B}\ }\textbf {\bibinfo {volume} {102}},\ \bibinfo
  {pages} {014307} (\bibinfo {year} {2020})}\BibitemShut {NoStop}%
\bibitem [{\citenamefont {Castro}\ \emph {et~al.}(2022)\citenamefont {Castro},
  \citenamefont {De~Giovannini}, \citenamefont {Sato}, \citenamefont
  {H\"ubener},\ and\ \citenamefont {Rubio}}]{Castro}%
  \BibitemOpen
  \bibfield  {author} {\bibinfo {author} {\bibfnamefont {A.}~\bibnamefont
  {Castro}}, \bibinfo {author} {\bibfnamefont {U.}~\bibnamefont
  {De~Giovannini}}, \bibinfo {author} {\bibfnamefont {S.~A.}\ \bibnamefont
  {Sato}}, \bibinfo {author} {\bibfnamefont {H.}~\bibnamefont {H\"ubener}},\
  and\ \bibinfo {author} {\bibfnamefont {A.}~\bibnamefont {Rubio}},\ }\bibfield
   {title} {\bibinfo {title} {Floquet engineering the band structure of
  materials with optimal control theory},\ }\href
  {https://doi.org/10.1103/PhysRevResearch.4.033213} {\bibfield  {journal}
  {\bibinfo  {journal} {Phys. Rev. Res.}\ }\textbf {\bibinfo {volume} {4}},\
  \bibinfo {pages} {033213} (\bibinfo {year} {2022})}\BibitemShut {NoStop}%
\bibitem [{\citenamefont {Higashikawa}\ \emph {et~al.}(2019)\citenamefont
  {Higashikawa}, \citenamefont {Nakagawa},\ and\ \citenamefont
  {Ueda}}]{Higashikawa}%
  \BibitemOpen
  \bibfield  {author} {\bibinfo {author} {\bibfnamefont {S.}~\bibnamefont
  {Higashikawa}}, \bibinfo {author} {\bibfnamefont {M.}~\bibnamefont
  {Nakagawa}},\ and\ \bibinfo {author} {\bibfnamefont {M.}~\bibnamefont
  {Ueda}},\ }\bibfield  {title} {\bibinfo {title} {Floquet chiral magnetic
  effect},\ }\href {https://doi.org/10.1103/PhysRevLett.123.066403} {\bibfield
  {journal} {\bibinfo  {journal} {Phys. Rev. Lett.}\ }\textbf {\bibinfo
  {volume} {123}},\ \bibinfo {pages} {066403} (\bibinfo {year}
  {2019})}\BibitemShut {NoStop}%
\bibitem [{\citenamefont {Li}\ \emph {et~al.}(2018)\citenamefont {Li},
  \citenamefont {Lee},\ and\ \citenamefont {Gong}}]{LiLinhu}%
  \BibitemOpen
  \bibfield  {author} {\bibinfo {author} {\bibfnamefont {L.}~\bibnamefont
  {Li}}, \bibinfo {author} {\bibfnamefont {C.~H.}\ \bibnamefont {Lee}},\ and\
  \bibinfo {author} {\bibfnamefont {J.}~\bibnamefont {Gong}},\ }\bibfield
  {title} {\bibinfo {title} {Realistic floquet semimetal with exotic
  topological linkages between arbitrarily many nodal loops},\ }\href
  {https://doi.org/10.1103/PhysRevLett.121.036401} {\bibfield  {journal}
  {\bibinfo  {journal} {Phys. Rev. Lett.}\ }\textbf {\bibinfo {volume} {121}},\
  \bibinfo {pages} {036401} (\bibinfo {year} {2018})}\BibitemShut {NoStop}%
\bibitem [{\citenamefont {Shirley}(1965)}]{JonHS}%
  \BibitemOpen
  \bibfield  {author} {\bibinfo {author} {\bibfnamefont {J.~H.}\ \bibnamefont
  {Shirley}},\ }\bibfield  {title} {\bibinfo {title} {{Solution of the
  Schrödinger Equation with a Hamiltonian Periodic in Time}},\ }\href
  {https://doi.org/10.1103/PhysRev.138.B979} {\bibfield  {journal} {\bibinfo
  {journal} {Phys. Rev.}\ }\textbf {\bibinfo {volume} {138}},\ \bibinfo {pages}
  {B979} (\bibinfo {year} {1965})}\BibitemShut {NoStop}%
\bibitem [{\citenamefont {Sambe}(1973)}]{HideoS}%
  \BibitemOpen
  \bibfield  {author} {\bibinfo {author} {\bibfnamefont {H.}~\bibnamefont
  {Sambe}},\ }\bibfield  {title} {\bibinfo {title} {{Steady States and
  Quasienergies of a Quantum-Mechanical System in an Oscillating Field}},\
  }\href {https://doi.org/10.1103/PhysRevA.7.2203} {\bibfield  {journal}
  {\bibinfo  {journal} {Phys. Rev. A}\ }\textbf {\bibinfo {volume} {7}},\
  \bibinfo {pages} {2203} (\bibinfo {year} {1973})}\BibitemShut {NoStop}%
\bibitem [{\citenamefont {Li}\ \emph {et~al.}(2025{\natexlab{b}})\citenamefont
  {Li}, \citenamefont {Cui}, \citenamefont {Zhang}, \citenamefont {Yu},\ and\
  \citenamefont {Yao}}]{LLeiSciBull}%
  \BibitemOpen
  \bibfield  {author} {\bibinfo {author} {\bibfnamefont {L.}~\bibnamefont
  {Li}}, \bibinfo {author} {\bibfnamefont {C.}~\bibnamefont {Cui}}, \bibinfo
  {author} {\bibfnamefont {R.-W.}\ \bibnamefont {Zhang}}, \bibinfo {author}
  {\bibfnamefont {Z.-M.}\ \bibnamefont {Yu}},\ and\ \bibinfo {author}
  {\bibfnamefont {Y.}~\bibnamefont {Yao}},\ }\bibfield  {title} {\bibinfo
  {title} {{Planar Hall plateau in magnetic Weyl semimetals}},\ }\href
  {https://doi.org/10.1016/j.scib.2024.11.026} {\bibfield  {journal} {\bibinfo
  {journal} {Sci. Bull.}\ }\textbf {\bibinfo {volume} {70}},\ \bibinfo {pages}
  {187} (\bibinfo {year} {2025}{\natexlab{b}})}\BibitemShut {NoStop}%
\bibitem [{\citenamefont {Li}\ \emph {et~al.}(2023)\citenamefont {Li},
  \citenamefont {Cao}, \citenamefont {Cui}, \citenamefont {Yu},\ and\
  \citenamefont {Yao}}]{LiLei}%
  \BibitemOpen
  \bibfield  {author} {\bibinfo {author} {\bibfnamefont {L.}~\bibnamefont
  {Li}}, \bibinfo {author} {\bibfnamefont {J.}~\bibnamefont {Cao}}, \bibinfo
  {author} {\bibfnamefont {C.}~\bibnamefont {Cui}}, \bibinfo {author}
  {\bibfnamefont {Z.-M.}\ \bibnamefont {Yu}},\ and\ \bibinfo {author}
  {\bibfnamefont {Y.}~\bibnamefont {Yao}},\ }\bibfield  {title} {\bibinfo
  {title} {{Planar Hall effect in topological Weyl and nodal-line
  semimetals}},\ }\href {https://doi.org/10.1103/PhysRevB.108.085120}
  {\bibfield  {journal} {\bibinfo  {journal} {Phys. Rev. B}\ }\textbf {\bibinfo
  {volume} {108}},\ \bibinfo {pages} {085120} (\bibinfo {year}
  {2023})}\BibitemShut {NoStop}%
\end{thebibliography}%
\end{document}